\pdfoutput=1
\PassOptionsToPackage{unicode}{hyperref}
\PassOptionsToPackage{hyphens}{url}
\PassOptionsToPackage{dvipsnames,svgnames,x11names}{xcolor}
\documentclass[12pt]{article}

\usepackage{amsmath,amssymb,amsthm,bm}
\usepackage{algorithm}
\usepackage{algpseudocode}

\usepackage{iftex}
\usepackage[T1]{fontenc}
\usepackage[utf8]{inputenc}
\usepackage{textcomp}
\usepackage{lmodern}
\IfFileExists{upquote.sty}{\usepackage{upquote}}{}
\IfFileExists{microtype.sty}{%
  \usepackage[]{microtype}
  \UseMicrotypeSet[protrusion]{basicmath}
}{}
\makeatletter
\@ifundefined{KOMAClassName}{%
  \IfFileExists{parskip.sty}{\usepackage{parskip}}{%
    \setlength{\parindent}{0pt}
    \setlength{\parskip}{6pt plus 2pt minus 1pt}}
}{\KOMAoptions{parskip=half}}
\makeatother
\usepackage{xcolor}
\makeatletter
\ifx\paragraph\undefined\else
  \let\oldparagraph\paragraph
  \renewcommand{\paragraph}{%
    \@ifstar\xxxParagraphStar\xxxParagraphNoStar}
  \newcommand{\xxxParagraphStar}[1]{\oldparagraph*{#1}\mbox{}}
  \newcommand{\xxxParagraphNoStar}[1]{\oldparagraph{#1}\mbox{}}
\fi
\ifx\subparagraph\undefined\else
  \let\oldsubparagraph\subparagraph
  \renewcommand{\subparagraph}{%
    \@ifstar\xxxSubParagraphStar\xxxSubParagraphNoStar}
  \newcommand{\xxxSubParagraphStar}[1]{\oldsubparagraph*{#1}\mbox{}}
  \newcommand{\xxxSubParagraphNoStar}[1]{\oldsubparagraph{#1}\mbox{}}
\fi
\makeatother

\usepackage{longtable,booktabs,array}
\usepackage{calc}
\usepackage{etoolbox}
\makeatletter
\patchcmd\longtable{\par}{\if@noskipsec\mbox{}\fi\par}{}{}
\makeatother
\IfFileExists{footnotehyper.sty}{\usepackage{footnotehyper}}{\usepackage{footnote}}
\makesavenoteenv{longtable}
\usepackage{graphicx}
\makeatletter
\def\maxwidth{\ifdim\Gin@nat@width>\linewidth\linewidth\else\Gin@nat@width\fi}
\def\maxheight{\ifdim\Gin@nat@height>\textheight\textheight\else\Gin@nat@height\fi}
\makeatother
\setkeys{Gin}{width=\maxwidth,height=\textheight,keepaspectratio}
\graphicspath{{Fig/}}
\makeatletter
\def\fps@figure{htbp}
\makeatother

\makeatletter
\@ifpackageloaded{caption}{}{\usepackage{caption}}
\AtBeginDocument{%
\ifdefined\contentsname\renewcommand*\contentsname{Table of contents}\else\newcommand\contentsname{Table of contents}\fi
\ifdefined\listfigurename\renewcommand*\listfigurename{List of Figures}\else\newcommand\listfigurename{List of Figures}\fi
\ifdefined\listtablename\renewcommand*\listtablename{List of Tables}\else\newcommand\listtablename{List of Tables}\fi
\ifdefined\figurename\renewcommand*\figurename{Figure}\else\newcommand\figurename{Figure}\fi
\ifdefined\tablename\renewcommand*\tablename{Table}\else\newcommand\tablename{Table}\fi}
\@ifpackageloaded{float}{}{\usepackage{float}}
\floatstyle{ruled}
\@ifundefined{c@chapter}{\newfloat{codelisting}{h}{lop}}{\newfloat{codelisting}{h}{lop}[chapter]}
\floatname{codelisting}{Listing}

\makeatother
\makeatletter
\@ifpackageloaded{caption}{}{\usepackage{caption}}
\@ifpackageloaded{subcaption}{}{\usepackage{subcaption}}
\makeatother

\ifLuaTeX
  \usepackage{selnolig}
\fi
\usepackage[authoryear,round]{natbib}
\usepackage{hyperref}
\usepackage{bookmark}
\IfFileExists{xurl.sty}{\usepackage{xurl}}{}
\hypersetup{
  pdftitle={Label Noise Cleaning for Supervised Classification via Bernoulli Random Sampling},
  pdfauthor={Yuxin Liu; Xiong Jin; Yang Han},
  pdfkeywords={Bernoulli random sampling, label noise, supervised classification, distribution of label noise level},
  colorlinks=true,
  linkcolor={blue},
  filecolor={Maroon},
  citecolor={Blue},
  urlcolor={Blue},
  pdfcreator={LaTeX via pandoc}}

\newcommand{\anon}{1}

\numberwithin{equation}{section}
\theoremstyle{plain}

\newtheorem{lemma}{Lemma}[section]

\newtheorem{corollary}{Corollary}[section]
\newtheorem{assumption}{Assumption}[section]
\theoremstyle{definition}
\newtheorem{definition}{Definition}[section]

\begin{document}

\def\spacingset#1{\renewcommand{\baselinestretch}{#1}\small\normalsize}
\spacingset{1}

\if1\anon
{
  \title{\bf Label Noise Cleaning for Supervised Classification via Bernoulli Random Sampling}
  \author{Yuxin Liu, Xiong Jin, Yang Han\thanks{Corresponding author. Department of Mathematics, The University of Manchester, Oxford Road, Manchester, M13 9PL, UK. E-mail: \href{mailto:yang.han@manchester.ac.uk}{yang.han@manchester.ac.uk}.}\\
       Department of Mathematics,  The University of Manchester, UK\\
}
 \maketitle
}
\fi

\if0\anon
{
  \bigskip
  \bigskip
  \bigskip
  \begin{center}
    {\LARGE\bf Label Noise Cleaning for Supervised Classification via Bernoulli Random Sampling}
  \end{center}
  \medskip
}
\fi

\bigskip
\begin{abstract}
Label noise - incorrect labels assigned to observations - can substantially degrade the performance of supervised classifiers. This paper proposes a label noise cleaning method based on Bernoulli random sampling. We show that the mean label noise levels of subsets generated by Bernoulli random sampling containing a given observation are identically distributed for all clean observations, and identically distributed, with a different distribution, for all noisy observations. Although the mean label noise levels are not independent across observations, by introducing an independent coupling we further prove that they converge to a mixture of two well-separated distributions corresponding to clean and noisy observations. By establishing a linear model between cross-validated classification errors and  label noise levels, we are able to approximate this mixture distribution and thereby separate clean and noisy observations without any prior label information. The proposed method is classifier-agnostic, theoretically justified, and demonstrates strong performance on both simulated and real datasets.
\end{abstract}

\vfill

\newpage
\spacingset{1.8} 

\section{Introduction}
In supervised classification, the objective is to use a labeled training dataset to learn a classifier that estimates the underlying mapping from the feature space to the label space. The trained classifier can accurately assign labels to new, unlabeled observations when the underlying mapping in the training dataset is consistent with the data to be classified \citep{han2020survey}. In practice, however, label noise, referring to incorrect labels on training observations, occurs due to human annotation errors or subjective biases \citep{han2020survey,jiang2020beyond}. 
Label noise can distort the mapping between features and labels, thereby reducing the performance of supervised classification \citep{zhu2004class,nettleton2010study,cannings2020classification}.
Therefore, cleaning label noise has become an important task in supervised classification problems \citep{wu2021ngc}.

Existing label noise cleaning methods can be broadly divided into two categories: 
algorithm-level methods and data-level methods \citep{johnson2022survey}.  
Algorithm-level methods focus on enhancing the robustness of classifiers on noisy training data \citep{ekambaram2016active, wang2018iterative, peterson2019human} or on employing alternative loss functions \citep{ghosh2017robust,han2018co} without modifying the training observations.
These methods are classifier-specific and may not generalize well to other classifiers.
  
In contrast, data-level methods aim to improve the quality of the training data and can be applied across various types of supervised classifiers. A commonly used approach in data-level methods is sub-sampling, which repeatedly draws subsets of fixed size, uses them to train supervised classifiers, and then evaluates the classifiers on the same or complementary subsets. Observations that are misclassified by one or more classifiers across these subsets are then regarded as potentially noisy \citep{brodley1996identifying, brodley1999identifying, frenay2013classification, garcia2019enabling}. Nevertheless, the theoretical basis for this identification remains limited, because the classifier’s model structure is not explicitly considered or when multiple classifiers are applied simultaneously.

In contrast to traditional sub-sampling methods, this paper uses Bernoulli random sampling to generate random subsets of the training data. Each observation defines a collection of subsets in which that observation is included. We prove that the mean label noise levels within the clean subgroups are identically distributed, and those within the noisy subgroups are identically distributed with a different distribution. However, these means are not independent across observations; rather, they are positively quadrant dependent. To address this, we introduce an independent coupling for each sequence. Each sequence is proved to converge to their independent coupling, which enables us to apply the strong law of large numbers and show that the empirical distribution of the mean label noise levels converges to a mixture distribution consisting of two well-separated sub-distributions corresponding to the clean and noisy groups.

We further show that, under specific conditions and assumptions, a linear model can be established between the label noise levels and the averaged cross-validation classification errors. This linear model allows us to approximate the mixture distribution of label noise levels. Since the mixture distribution is shown to be well separated, the midpoint between the expectations of the two sub-distributions provides a natural cut-off for separating clean from noisy observations. Moreover, by matching the ratio of density areas to the ratio of observations, we can identify an approximate cut-off among the averaged cross-validation errors without explicitly estimating the linear model parameters. To enhance stability from the approximation errors, we further propose a stepwise procedure that iteratively refines the separation until a self-consistency criterion is satisfied.

Some studies have also explored label noise cleaning by modeling the distribution of supervised classification performance in datasets with label noise under pre-specified distributional assumptions. For example, \citet{arazo2019unsupervised} assumed that this distribution follows a Beta Mixture Model (BMM), validated the assumption through simulations, and developed an EM-based procedure for estimating the model parameters.
In contrast, our method does not rely on any pre-assumed distributional form. We show that the distribution of supervised classification errors approximates a mixture of sub-distributions with explicit characterizations, without requiring any prior label information.
 
The structure of this paper is as follows.
Section \ref{sec2} develops the distributional theory underlying our method. It shows that the clean and noisy observations are each  identically distributed. It then shows the positive quadrant dependence among observations and their convergence to an independent coupling. Finally, it gives the resulting mixture distribution over all observations.
Section \ref{sec3} introduces the linear model between label noise levels and averaged cross-validation errors and shows how an error-based cut-off enables approximate clean/noisy observations separation without label information. A stepwise refinement and the complete algorithm are also presented.
Section \ref{sec4} gives experimental results on both simulated and real datasets and compares our method with existing approaches.
Section \ref{sec5} concludes the paper.

\section{Problem setup and theoretical basis for separation}
\label{sec2}
\subsection{Dataset with label noise, supervised classification, and assumptions}

\begin{definition}[]\label{def:mixture}
Suppose the feature space is $\mathbb{R}^p$ $(p \ge 1)$, meaning each observation has a $p$-dimensional real-valued feature vector, and the label space consists of $K$ discrete classes denoted by $[K] := \{1,2,\dots,K\}$. Define the orthogonal projections
\[
\Pi_{\mathbb{R}^p} : \mathbb{R}^p \times [K] \to \mathbb{R}^p,
\qquad 
\Pi_{[K]} : \mathbb{R}^p \times [K] \to [K],
\]
by
\[
\Pi_{\mathbb{R}^p}(\mathbf{x}, y) = \mathbf{x}, 
\qquad 
\Pi_{[K]}(\mathbf{x}, y) = y.
\]
Let $P_1$ and $P_2$ be two probability distributions on $\mathbb{R}^p \times [K]$ such that  
\[
\Pi^{*}_{\mathbb{R}^p}(P_1) = \Pi^{*}_{\mathbb{R}^p}(P_2),
\qquad
\Pi^{*}_{[K]}(P_1) \neq \Pi^{*}_{[K]}(P_2),\qquad\text{where $\Pi^{*}(P) := P \circ \Pi^{-1}$.}
\]
That is, $P_1$ and $P_2$ share the same marginal distribution on the feature space $\mathbb{R}^p$ but differ in their label marginals on $[K]$.  

Define the mixture distribution
\[
P = \pi_1 P_1 + \pi_2 P_2,
\qquad 
\pi_1 + \pi_2 = 1, \quad \pi_1 > \pi_2 > 0.
\]
Thus, a sample drawn from $P$ comes from $P_1$ (the clean sub-distribution) with probability $\pi_1$ and from $P_2$ (the noisy sub-distribution) with probability $\pi_2$.

Let
\[
D = \{\bm{z}_i = (\bm{x}_i, y_i)\}_{i=1}^{N}
\]
denote the observed dataset, whose samples are drawn independently from the mixture distribution $P$.  
The dataset $D$ can be decomposed into two disjoint subsets,
\[
D = D_1 \cup D_2,
\qquad
D_1 \cap D_2 = \emptyset,
\]
where
\begin{itemize}
    \item $D_1 := \{(\bm{x}_i, y_i) \in D : (\bm{x}_i, y_i) \sim P_1\}$ corresponds to observations generated from $P_1$, and
    \item $D_2 := \{(\bm{x}_i, y_i) \in D : (\bm{x}_i, y_i) \sim P_2\}$ corresponds to observations generated from $P_2$.
\end{itemize}

As $N \to \infty$, by the strong law of large numbers, the empirical proportions of these subsets converge to their theoretical mixture probabilities:
\[
\frac{N_1}{N} \xrightarrow{p} \pi_1,
\qquad
\frac{N_2}{N} \xrightarrow{p} \pi_2,
\]
\end{definition}
where $N_1$ is the size of $D_1$, $N_2$ is the size of $D_2$.

A supervised classification problem with noisy training data consists of training a classifier $h$ on $D$, denoted $h_D$, where
\[
h_D : \mathbb{R}^p \;\to\; [K].
\]

For test data $D^{\mathrm{te}} = \{\bm{z}_i^{\mathrm{te}}\}_{i=1}^{N^{\mathrm{te}}}$, 
the error rate of $h_D$ is
\[
e({h_{D}};D^{\mathrm{te}})
= \frac{1}{N^{\mathrm{te}}}\sum_{i=1}^{N^{\mathrm{te}}}
   \mathbb{I}\{\,h_D(\bm{x}_i^{\mathrm{te}})\neq y_i^{\mathrm{te}}\,\}.
\]
For the test error $e(h_{D};D^{\mathrm{te}})$, we impose the following two assumptions throughout this paper.

\begin{assumption}[Ergodicity]\label{ass:ergodicity}
For sufficiently large $N$ and $N^{\mathrm{te}}$, the empirical test error converges almost surely to its population counterpart:
\[
e(h_{D}; D^{\mathrm{te}}) \xrightarrow{a.s.} e(h_{P}; P^{\mathrm{te}}),
\]
where $P^{\mathrm{te}} = \pi^{\mathrm{te}}_1 P_1 + \pi^{\mathrm{te}}_2 P_2$ denotes the underlying mixture distribution of the test data, with $\pi^{\mathrm{te}}_1 + \pi^{\mathrm{te}}_2 = 1$.  

Here, $e(h_{P}; P^{\mathrm{te}})$ represents the population-level generalization error of the classifier $h_P$ trained on the population distribution $P = \pi_1 P_1 + \pi_2 P_2$ and evaluated on the test distribution $P^{\mathrm{te}}$.  
Formally, it is defined as
\[
e(h_{P}; P^{\mathrm{te}}) := \mathbb{E}_{(\bm{x}, y) \sim P^{\mathrm{te}}}\!\left[\mathbb{I}\{h_{P}(\bm{x}) \neq y\}\right],
\]
that is, the expected misclassification probability of $h_P$ under the test distribution $P^{\mathrm{te}}$.
\end{assumption}

\begin{assumption}[Monotonicity]\label{ass:monotonicity}
If $\pi_1 > \pi_2 > 0$ and $\pi^{\mathrm{te}}_1 > \pi^{\mathrm{te}}_2$,  
then $e(h_{P}; P_{\mathrm{te}}) \in (0,1)$  
is a monotonically increasing function of both $\pi_2$ and $\pi^{\mathrm{te}}_2$.
\end{assumption}

The ergodicity assumption is justified because, under common conditions such as i.i.d.\ sampling and stable learning procedures, the empirical test error naturally converges to its population value as the dataset grows.  
The monotonicity assumption reflects a natural property of classification under label noise: when the number of observations is sufficiently large in a dataset, as the proportion of clean samples increases in either training or test data, the classifier is exposed to fewer mislabeled examples, leading to a lower expected error.

\subsection{A positively quadrant dependent (PQD) sequence of random variables 
\texorpdfstring{$\{\tilde{l}_{\bm{z}}\}_{\bm{z}\in D}$}{\{lz\}} and its separated empirical distribution}

\label{subsec2}
Let $M \ge 1$ denote the number of Bernoulli random subsets to be sampled. 
Define 
\[
\{\xi_{\bm{z},m} : \bm{z} \in D,\ 1 \le m \le M\}
\]
as a sequence of i.i.d.\ random variables following the Bernoulli distribution $\mathrm{Be}(q)$. 
For each $1 \le m \le M$, let 
\[
B_m = \{\bm{z} \in D : \xi_{\bm{z},m} = 1\}.
\]
$B_m$ is referred to as a Bernoulli random subset of $D$, and its law is denoted by $\mathrm{BS}_q(D)$.

For any subset $B \subset D$, we define its \emph{label noise level} as follows.

\begin{definition}[Label Noise Level]\label{def:label noise level}
The label noise level of a subset $B \subset D$ is defined by
\[
l(B) = \frac{\#(B \cap D_2)}{\# D}.
\]
\end{definition}

Given $\bm{z} \in D$, for $1 \le m \le M$, the random variable $l(B_m)$ conditional on $\bm{z} \in B_m$, denoted by $l(B_m \mid \bm{z} \in B_m)$, can be equivalently written as $l(B_m^{(\bm{z})} \cup \{\bm{z}\})$, where 
\[
B_m^{(\bm{z})} = B_m \setminus \{\bm{z}\} = \{\bm{z}' \in D \setminus \{\bm{z}\} : \xi_{\bm{z}',m} = 1\}
\]
is independent of $\{\xi_{\bm{z},m}\}_{1 \le m \le M}$.

For each $\bm{z} \in D$, define  
\[
\mathcal{I}_{\bm{z}} = \{1 \le m \le M : \xi_{\bm{z}, m} = 1\}, \qquad 
\# \mathcal{I}_{\bm{z}} = \sum_{m=1}^{M} \xi_{\bm{z}, m}
\]
and 
\[
\tilde{l}_{\bm{z}} = \frac{\sum_{m \in \mathcal{I}_{\bm{z}}} l(B_m^{(\bm{z})} \cup \{\bm{z}\})}{\# \mathcal{I}_{\bm{z}}}.
\]
The random variables $\{l(B_m^{(\bm{z})}\cup\{\bm{z}\})\}_{m \in \mathcal{I}_{\bm{z}}}$ are independent of $\#\mathcal{I}_{\bm{z}}$, $\#\mathcal{I}_{\bm{z}}\sim Bin(M,q)$.

Let $l_i$ denote the distribution of 
\[
\{\,l(B_m^{(\bm{z})}\cup\{\bm{z}\}) : \bm{z}\in D_i\,\}_{m\in\mathcal{I}_{\bm{z}}}, \qquad i=1,2.
\]
For each $i$, 
$\{\,l(B_m^{(\bm{z})}\cup\{\bm{z}\}) : \bm{z}\in D_i\,\}_{m\in\mathcal{I}_{\bm{z}}}$
forms a sequence of identically distributed random variables with distribution $l_i$.

Consequently, for any $\bm{z}\in D_i$, the empirical average $\tilde{l}_{\bm{z}}$ has distribution
\[
L_i \;=\; \frac{1}{T}\sum_{j=1}^{T} l_i^{(j)}, \qquad i=1,2,
\]
where $T \sim \mathrm{Bin}(M,q)$ and $l_i^{(j)} \stackrel{\mathrm{i.i.d.}}{\sim} l_i$ for $j=1,\ldots,T$, with $T$ independent of $\{l_i^{(j)}\}_{j=1}^{T}$.

Lemma~\ref{lem:2.1} provides the explicit expressions of $l_1$ and $l_2$, and shows that $L_1$ and $L_2$ differ by a fixed mean gap.

\begin{lemma}\label{lem:2.1}
The distributions $l_1$ and $l_2$ are given by
\[
l_1 \;\sim\; \frac{T_2 + \xi}{\,T_1 + T_2 + \xi + 1\,}, 
\qquad
l_2 \;\sim\; \frac{T_2 + 1}{\,T_1 + T_2 + \xi + 1\,},
\]
where $T_1 \sim \mathrm{Bin}(N_1 - 1, q)$, $T_2 \sim \mathrm{Bin}(N_2 - 1, q)$, and 
$\xi \sim \mathrm{Be}(q)$, with $\xi$, $T_1$, and $T_2$ mutually independent.

Let $\mu_1 = \mathbb{E}[L_1]$ and $\mu_2 = \mathbb{E}[L_2]$ denote the expectations of $L_1$ and $L_2$, then
\[
\mu_1 = \mathbb{E}[l_1]=\mathbb{E}[\frac{T_2+\xi}{T_1+T_2+\xi+1}],\qquad \mu_2 = \mathbb{E}[\frac{T_2+1}{T_1+T_2+\xi+1}]
\]
and
\[
\mu_2 - \mu_1 
= \mathbb{E}\!\left[\frac{1-\xi}{T_1+T_2+\xi+1}\right]
\approx \frac{1 - q}{(N - 1)q + 1} > 0.
\]
\end{lemma}

Since every individual component among $\{\tilde{l}_{\bm{z}}\}_{\bm{z}\in D_1}$ identically follows $L_1$, and every individual component among $\{\tilde{l}_{\bm{z}}\}_{\bm{z}\in D_2}$ identically follows $L_2$. If the Strong Law of Large Numbers (SLLN) holds, then the empirical distribution of $\{\tilde{l}_{\bm{z}}\}_{\bm{z}\in D}$ over $\bm{z}$ will converge to the mixture of $L_1$ and $L_2$. By Lemma~\ref{lem:2.1}, since the two distributions have a fixed mean gap. As $M\to\infty$, the mixture separates into two distinct distributions, and therefore $D_1$ and $D_2$ can be separated.  

However, although the components of $\{\tilde{l}_{\bm{z}}\}_{\bm{z}\in D_1}$ and $\{\tilde{l}_{\bm{z}}\}_{\bm{z}\in D_2}$ are respectively identically distributed, they are not independent; instead, they are positively quadrant dependent (PQD). The definition of PQD is given in Definition~\ref{def:pqd}, and this dependence structure is stated in Lemma~\ref{lem:2.2}.

\begin{definition}[PQD Random Variables]\label{def:pqd}
A pair of random variables $(X, Y)$ is said to be \emph{PQD} if, for all $x, y \in \mathbb{R}$,
\[
P(X \le x,\, Y \le y) \;\ge\; P(X \le x)\, P(Y \le y).
\]
\end{definition}

\begin{lemma}\label{lem:2.2}
$\{\tilde{l}_{\bm{z}}\}_{\bm{z}\in D_1}$ and $\{\tilde{l}_{\bm{z}}\}_{\bm{z}\in D_2}$  
each forms a sequence of pairwise PQD random variables.  
\end{lemma}


Strong laws of large numbers (SLLN) for PQD random variables have been established in the literature (see, e.g., \citep{birkel1988note}).  
However, such results require a correlation decay condition 
\[
\sum_{j=1}^{N} \frac{1}{j^2} \operatorname{Cov}\!\left(\tilde{l}_{\bm{z}_j}, \sum_{i=1}^{j} \tilde{l}_{\bm{z}_i}\right) < \infty,
\]
so that the SLLN holds for the sequence $\{\tilde{l}_{\bm{z}}\}_{\bm{z}\in D}$.  
In our setting, the correlation decay behavior is unclear. In addition, the variables $\{\tilde{l}_{\bm{z}}\}_{\bm{z}\in D}$ are not drawn from a common distribution that is independent of $N$; rather, each $\tilde{l}_{\bm{z}}$ depends on $N$.  
Hence the SLLN for PQD random variables is not applicable in this case.

In Section~\ref{sec2.3}, we show that the SLLN can be applied to an independent coupling of $\{\tilde{l}_{\bm{z}}\}_{\bm{z}\in D}$, and that the empirical distribution of $\{\tilde{l}_{\bm{z}}\}_{\bm{z}\in D}$ asymptotically converges to the empirical distribution of this independent coupling.

\subsection{An independent coupling and 
the approximate empirical distribution of \texorpdfstring{$\{\tilde{l}_{\bm{z}}\}_{\bm{z}\in D}$}{\{lz\}}}
\label{sec2.3}

Let $\{B'_{\bm{z},m}\}_{\bm{z}\in D,\, m=1,\ldots,M}$ be a collection of independent random subsets such that, for each $\bm{z} \in D$, the subset $B'_{\bm{z},m}$ follows the law $\mathrm{RS}_q(D \setminus \{\bm{z}\})$.  
We further assume that $\{B'_{\bm{z},m}\}_{\bm{z}\in D,\, m=1,\ldots,M}$ is independent of $\{\xi_{\bm{z},m} : \bm{z} \in D,\, 1 \le m \le M\}$.  
Define  
\[
\tilde{l}'_{\bm{z}} = \frac{\sum_{m \in \mathcal{I}_{\bm{z}}} l(B'_{\bm{z},m} \cup \{\bm{z}\})}{\# \mathcal{I}_{\bm{z}}}.
\]
Each $\tilde{l}'_{\bm{z}}$ is identically distributed with $\tilde{l}_{\bm{z}}$, and the sequence $\{\tilde{l}'_{\bm{z}}\}_{\bm{z}\in D}$ is independent, since $\{\mathcal{I}_{\bm{z}}\}_{\bm{z}\in D}$ are independent.  
We refer to $\{\tilde{l}'_{\bm{z}}\}_{\bm{z}\in D}$ as an independent coupling of $\{\tilde{l}_{\bm{z}}\}_{\bm{z}\in D}$.

Let $g : [0,1] \to \mathbb{R}_+$ be a differentiable function satisfying $\|g'\|_{\infty} = C_g < \infty$.  
The following lemma shows that the discrepancy
\[
\frac{1}{N} \sum_{\bm{z} \in D} \bigl( g(\tilde{l}_{\bm{z}}) - g(\tilde{l}'_{\bm{z}}) \bigr),
\]
converges to 0 as $M>N\to\infty$.

\begin{lemma}\label{lemma3}
When $M\to\infty$ and $N\to\infty$,
\[
\mathbb{E}\left(\sum_{\bm{z}\in D_{1}} g(\tilde{l}_{\bm{z}})-g(\tilde{l}'_{\bm{z}})\right)^2= O(\frac{N}{M}),\qquad \mathbb{E}\left(\sum_{\bm{z}\in D_{2}} g(\tilde{l}_{\bm{z}})-g(\tilde{l}'_{\bm{z}})\right)^2= O(\frac{N}{M}).
\]
\end{lemma}

Lemma~\ref{lemma3} shows the convergence of each subsequence to its respective independent coupling.
Building on this result, the following lemma extends the argument to the combined sequence ${\{\tilde{l}_{\bm{z}}\}}_{\bm{z}\in D}$, showing that it converges to its independent coupling ${\{\tilde{l}'_{\bm{z}}\}}_{\bm{z}\in D}$ without requiring identification of whether $\bm{z} \in D_1$ or $\bm{z} \in D_2$.

\begin{lemma}\label{lemma4}
Assume that $M\ge N^{\eta}$ for some $\eta>0$. Let $g:[0,1]\to \mathbb{R}_+$ be a differentiable function with $\|g'\|_{\infty}<\infty$. With probability $1$ there exists an integer $N_0<\infty$ such that for all $N\ge N_0$,
\[
\frac{1}{N}\left|\sum_{\bm{z}\in D} g(\tilde{l}_{\bm{z}})-g(\tilde{l}'_{\bm{z}})\right|< N^{-\frac{\eta}{3}}.
\]
\end{lemma}


One may define the space of test functions as  
\[
\mathcal{G} = \{\, g : [0,1] \to \mathbb{R}_+ \mid g \text{ is differentiable and } \|g'\|_{\infty} < \infty \,\}.
\]
By selecting a countable dense subset of $\mathcal{G}$ with respect to the uniform norm—and applying Lemma~\ref{lemma4},  
we deduce that the empirical distributions of $\{\tilde{l}_{\bm{z}}\}_{\bm{z}\in D}$ asymptotically converges to that of $\{\tilde{l}'_{\bm{z}}\}_{\bm{z}\in D}$  as $N \to \infty$.  
Since the components of $\{\tilde{l}'_{\bm{z}}\}_{\bm{z}\in D}$ are i.i.d., the SLLN can be applied to $\{\tilde{l}'_{\bm{z}}\}_{\bm{z}\in D}$. Moreover, the subsequences  
$\{\tilde{l}'_{\bm{z}}\}_{\bm{z}\in D_1}$ and $\{\tilde{l}'_{\bm{z}}\}_{\bm{z}\in D_2}$ have the same  mean gap as $\{\tilde{l}'_{\bm{z}}\}_{\bm{z}\in D_1}$ and $\{\tilde{l}'_{\bm{z}}\}_{\bm{z}\in D_2}$, as revealed in Lemma~\ref{lem:2.1}. Hence we have the following corollary.

\begin{corollary}\label{corollary1}
Assume that $M \ge N^{\eta}$ for some $\eta > 0$.  
Then, with probability 1, for sufficiently large $N$, the empirical average
\[
\frac{1}{N}\sum_{\bm{z} \in D} \delta_{\tilde{l}_{\bm{z}}}
\]
 asymptotically converges to the mixture distribution
\[
L \stackrel{\mathrm{a}}{\sim}
\begin{cases}
L_1, & \text{with probability } \pi_1, \\[0.8ex]
L_2, & \text{with probability } \pi_2,
\end{cases}
\]
where $L_1\stackrel{\mathrm{a}}{\sim} \mathcal{N}(\mu_1,\sigma_1^2)$, $L_2\stackrel{\mathrm{a}}{\sim} \mathcal{N}(\mu_2,\sigma_2^2)$, with $\sigma_1^2=O(\frac{1}{MN})$, $\sigma_2^2=O(\frac{1}{MN})$.
\end{corollary}

Corollary~\ref{corollary1} shows that, when $M \ge N^{\eta}$ for some $\eta > 0$ and $N \to \infty$, the empirical distribution of $\{\tilde{l}_{\bm{z}}\}_{\bm{z}\in D}$ is asymptotically a mixture of $L_1$ and $L_2$, corresponding respectively to $D_1$ and $D_2$.

\begin{figure}[t]
    \centering
        \includegraphics[width=11cm]{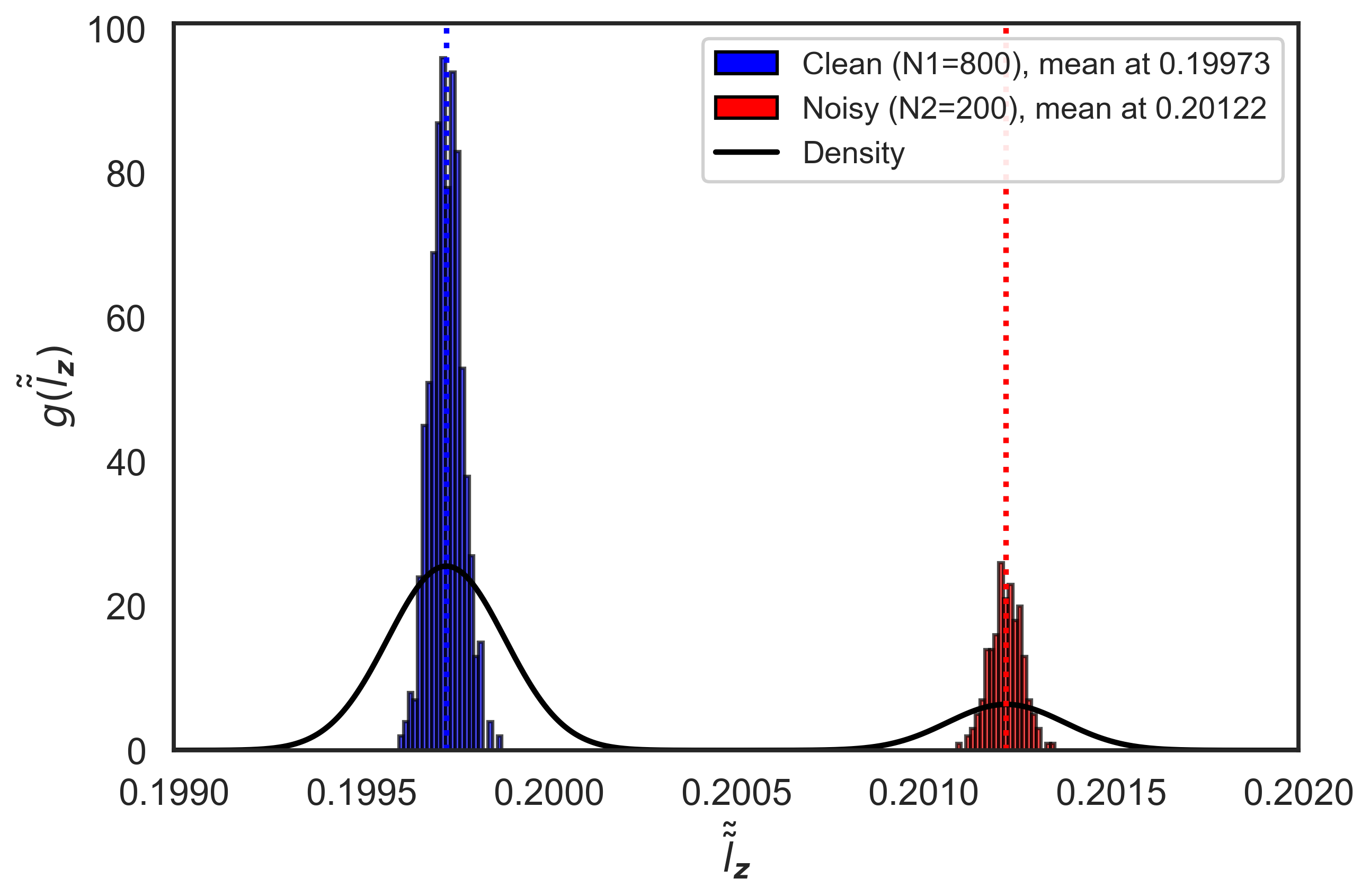}
   \caption{Empirical distribution of $\{\tilde{l}_{\bm{z}}\}_{\bm{z} \in D}$ obtained from a dataset with $N = 1000$ observations ($N_1 = 800$ clean and $N_2 = 200$ noisy).  
The values of $\tilde{l}_{\bm{z}}$ are computed under the Bernoulli random sampling scheme with $M = 2 \times 10^6$ subsets and inclusion probability $q = 0.4$.  
The histogram shows two well-separated laws corresponding to $L_1$ and $L_2$, with a mean gap
$\mathbb{E}\!\left[\tfrac{1 - q}{(N - 1)q + 1}\right] \approx 0.0015$.}
\label{fsupp1}
\end{figure}

Figure~\ref{fsupp1} presents an illustrative example demonstrating the mixed distribution of ${\{\tilde{l}_{\bm{z}}\}}_{\bm{z}\in D}$, obtained from a dataset $D$ of size $N = 1000$, consisting of $N_1 = 800$ clean and $N_2 = 200$ noisy observations.  
Under the Bernoulli random sampling framework, $M = 2 \times 10^6$ Bernoulli random subsets are generated with inclusion probability $q = 0.4$.  
For each observation $\bm{z}$, all subsets containing $\bm{z}$ are collected, and the average label noise level within this collection is used to compute $\tilde{l}_{\bm{z}}$.  
The empirical distribution of $\{\tilde{l}_{\bm{z}}\}_{\bm{z} \in D}$ is then approximated over all $\bm{z} \in D$ and shown in Figure~\ref{fsupp1}.  

The resulting distribution exhibits two clearly separated components corresponding to the clean subset $D_1$ and the noisy subset $D_2$.  
The difference between their component means closely aligns with the theoretical separation established in Lemma~\ref{lem:2.1}, given by  
$\mathbb{E}\!\left[\tfrac{1 - q}{(N - 1)q + 1}\right] \approx 0.0015$.

\begin{figure}[t]
    \centering
    \includegraphics[width=11cm]{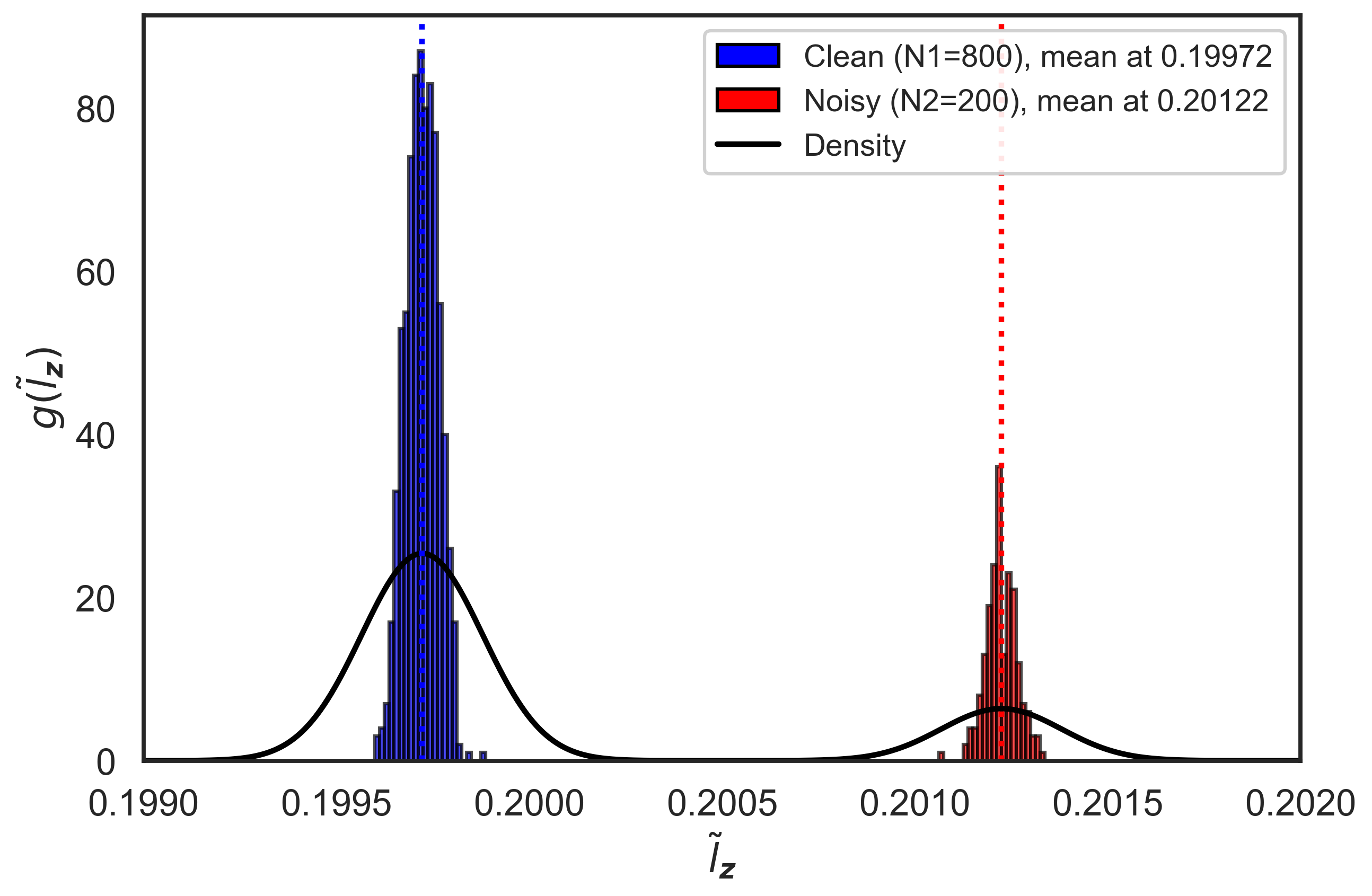}
\caption{Empirical distribution of the i.i.d.\ sequence $\{\tilde{l}'_{\bm{z}}\}_{\bm{z} \in D}$ generated under the same settings as Figure~\ref{fsupp1}.  
The mixture shows two components, approximately corresponding to $L_1$ and $L_2$, with a mean gap of $\tfrac{1 - q}{(N - 1)q + 1} \approx 0.0015$, consistent with Lemma~\ref{lemma3}.}

 \label{fsupp2}
\end{figure}

Figure~\ref{fsupp2} illustrates a complementary example constructed under the same parameter settings as in Figure~\ref{fsupp1}  
($N_1$, $N_2$, $M$, and $q$).  
In contrast to Figure~\ref{fsupp1}, here each observation $\bm{z} \in D$ is treated conditionally on its inclusion in $B$.  
Given $\bm{z} \in B$, we sample $M$ subsets $B^{(\bm{z})}$ according to the Bernoulli random sampling framework, and form $B = B^{(\bm{z})} \cup \{\bm{z}\}$.  
All such subsets are then used to compute $\tilde{l}'_{\bm{z}}$.  
Repeating this procedure over all $\bm{z} \in D$ yields the empirical distribution of $\{\tilde{l}'_{\bm{z}}\}_{\bm{z} \in D}$.  
As shown in Figure~\ref{fsupp2}, the resulting distribution exhibits a nearly identical two-component mixture structure as in Figure~\ref{fsupp1}, with the same mean separation  
$\tfrac{1 - q}{(N - 1)q + 1} \approx 0.0015$.  
This result empirically confirms the convergence as shown in Corollary~\ref{corollary1}.

In this section, we have established that the empirical distribution of $\{\tilde{l}_{\bm{z}}\}_{\bm{z}\in D}$ can be separated according to whether $\bm{z} \in D_1$ or $\bm{z} \in D_2$.  
In the next section, we extend this result by linking $\{\tilde{l}_{\bm{z}}\}_{\bm{z}\in D}$ to supervised classification.

\section{Separation of clean and noisy observations using averaged cross-validation supervised classification error}\label{sec3}
\subsection{Linear modeling of \texorpdfstring{$\{\tilde{l}_{\bm{z}}\}_{\bm{z}\in D}$}{\{lz\}} via averaged cross-validation error}\label{subsec3}

In practice, we do not know which observations are clean or noisy, so the true label noise level \(l(B)\) cannot be directly computed for any subset \(B \subset D\). 
We therefore use the averaged cross-validation error on a subset \(B\) as a surrogate for the unobservable label noise level \(l(B)\), which is formally defined in Definition~\ref{def:cv-error}.

\begin{definition}[]\label{def:cv-error}
Let \(B\) be a \emph{Bernoulli random subset} of a dataset \(D\), and let 
\(\{(B^{(j)}_{\mathrm{train}}, B^{(j)}_{\mathrm{test}})\}_{j=1}^{k}\) 
denote a \(k\)-fold random partition of \(B\), 
with $B = B^{(j)}_{\mathrm{train}} \cup B^{(j)}_{\mathrm{test}}$ and 
$B^{(j)}_{\mathrm{train}} \cap B^{(j)}_{\mathrm{test}} = \emptyset$.  
The averaged cross-validation error of $B$ is defined as
\[
e_{\mathrm{cv}}(h,B):= \frac{1}{k} \sum_{j=1}^{k} e(h_{B^{(j)}_{\mathrm{train}}};B^{(j)}_{\mathrm{test}}).
\]
\end{definition}

For each $\bm{z} \in D$, define
\[
\tilde{e}_{\bm{z}} = \frac{\sum_{m \in \mathcal{I}_{\bm{z}}} e_{\mathrm{cv}}(h, B_m^{(\bm{z})} \cup \{\bm{z}\})}{\# \mathcal{I}_{\bm{z}}}.
\]

Lemma~\ref{lem:linear} proves that, $\tilde{l}_{\bm{z}}$ can be well approximated by a linear model from $\tilde{e}_{\bm{z}}$.

\begin{lemma}\label{lem:linear}
Suppose $\pi_1>\pi_2 > 0$. For sufficiently large $N$ and $M$,  
the empirical statistics 
$\tilde{\bm{l}} = (\tilde{l}_{\bm{z}})_{\bm{z} \in D}$ and 
$\tilde{\bm{e}} = (\tilde{e}_{\bm{z}})_{\bm{z} \in D}$ satisfy the approximate linear relation
\[
\tilde{\bm{l}} = \beta\, \tilde{\bm{e}} + \bm{\epsilon},
\]
where the error term $\bm{\epsilon} = (\epsilon_{\bm{z}})_{\bm{z} \in D}$ is approximately distributed as  
$N\!\left(0,\, \tfrac{1}{qM}\, \Sigma\right)$,  
$\beta$ is a positive common scaling coefficient, and  
$\Sigma = (\sigma_{\bm{z}_i \bm{z}_j})_{i,j=1}^{N}$ denotes the covariance matrix.
\end{lemma}

\begin{figure}[H]
\centering
\includegraphics[width=14cm]{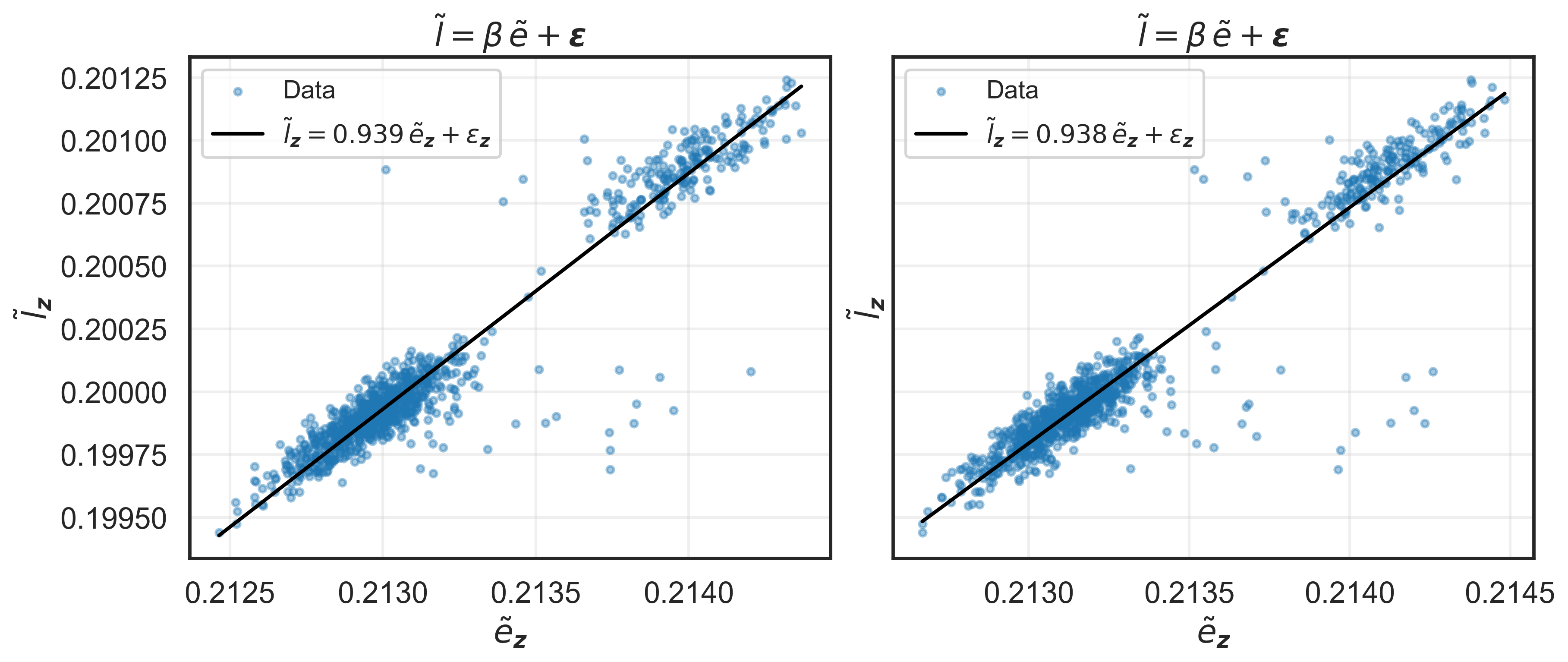}
\caption{Linear model between $\tilde{l}_{\bm{z}}$ and $\tilde{e}_{\bm{z}}$ under Setting~1 with 20\% label noise. 
Results are shown for classifiers trained with RBF SVM and 1-NN, as described in Section~\ref{sec4}. }
\label{f3}
\end{figure}

Figure~\ref{f3} is an illustrative example of this linear model using 1000 observations generated under simulated Data Setting~1 (see Section~\ref{sec4}) with 20\% label noise, evaluated with two different supervised classifiers, radial basis function SVM (RBF SVM) and 1-Nearest Neighbor (1-NN).

\subsection{Cut point selection for separating clean and noisy observations}

In Section~\ref{subsec2}, we established that the sequences $\{\tilde{l}_{\bm{z}}\}_{\bm{z}\in D_1}$ and $\{\tilde{l}_{\bm{z}}\}_{\bm{z}\in D_2}$ follow distinct distributions.  
Therefore, selecting the midpoint between their respective expectations provides a natural separation between the clean and noisy subsets, as formally defined in Definition~\ref{def:lstar}.

\begin{definition}[Cut point $l^{*}$]\label{def:lstar}
The \emph{cut point} $l^{*}$, which separates the clean and noisy subsets $D_1$ and $D_2$, is defined as the midpoint between their respective expected label noise levels:
\[
l^{*} 
:= \frac{1}{2}\!\left(
\mathbb{E}_{\bm{z}\sim P_1}[\tilde{l}_{\bm{z}}]
+
\mathbb{E}_{\bm{z}\sim P_2}[\tilde{l}_{\bm{z}}]
\right).
\]
\end{definition}

Based on the linear model introduced in Section~\ref{subsec3}, Lemma~\ref{lem:ecut} provides an explicit expression for the cut point in terms of $\{\tilde{e}_{\bm{z}}\}_{\bm{z}\in D}$, which yields an approximate separation between clean and noisy observations, without requiring the estimation of the linear coefficient $\beta$ or any values of $\{\tilde{l}_{\bm{z}}\}_{\bm{z}\in D}$.

\begin{lemma}[]\label{lem:ecut}
For sufficiently large $N$ and $M$, the cut point $e^{*}$ for separating clean and noisy observations in $D$ can be estimated as
\[
e^{*} 
\approx \mathop{\arg\min}_{c \in \left(\min\{\tilde{e}_{\bm{z}}\}_{\bm{z}\in D},\, \max\{\tilde{e}_{\bm{z}}\}_{\bm{z}\in D}\right)}
\left| \frac{\sum_{\bm{z} \in \{\tilde{e}_{\bm{z}} < c\}} (\tilde{e}_{\bm{z}} - c)}
           {\sum_{\bm{z} \in \{\tilde{e}_{\bm{z}} \geq c\}} (c - \tilde{e}_{\bm{z}})}
     - \frac{\sum_{\bm{z} \in D} \mathbb{I}\{\tilde{e}_{\bm{z}} < c\}}
            {\sum_{\bm{z} \in D} \mathbb{I}\{\tilde{e}_{\bm{z}} \geq c\}} \right|.
\]
\end{lemma}

\begin{figure}[H] 
\centering 
\includegraphics[width=14cm]{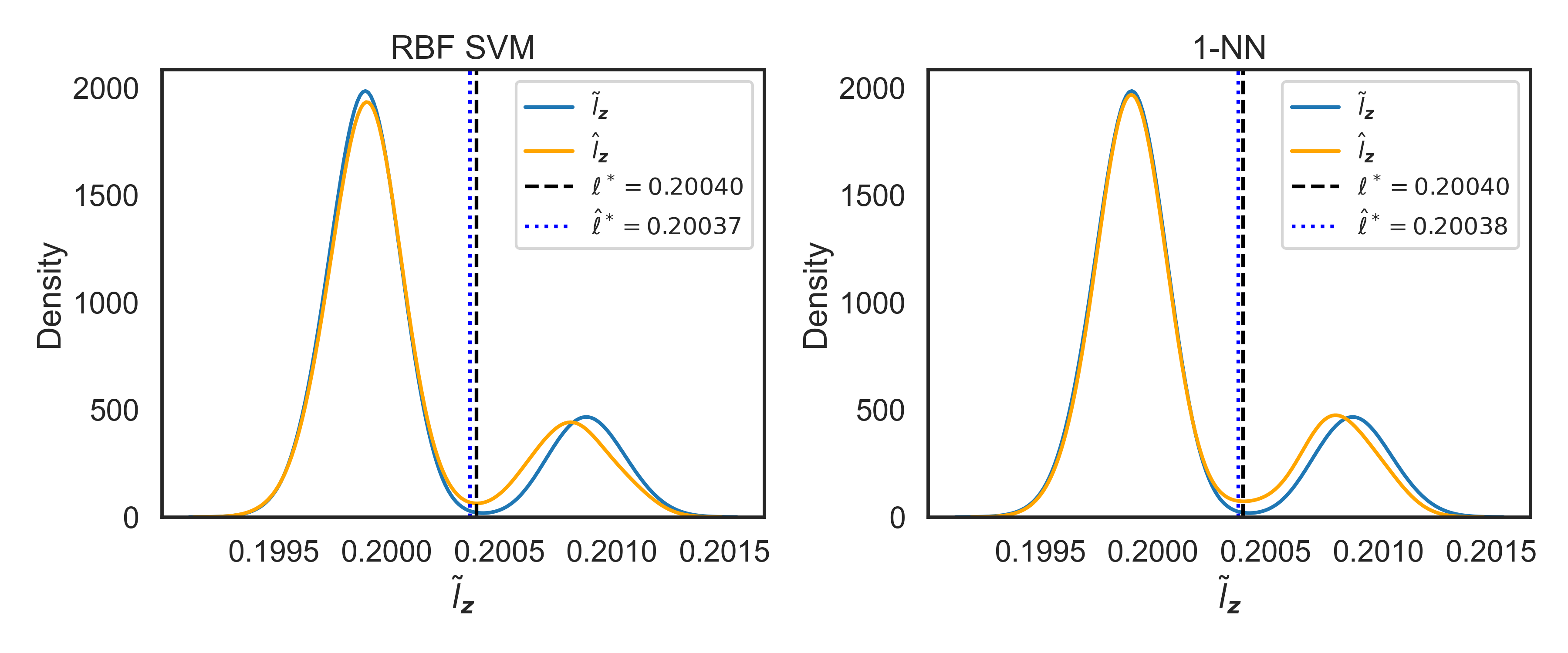} \caption{Comparison between the empirical and model-estimated distributions of $\tilde{l}_{\bm{z}}$.  
    The estimated distribution $\hat{l}_{\bm{z}}$ closely matches the empirical $\tilde{l}_{\bm{z}}$, 
    and the estimated cut point $\hat{l}^{*}$ aligns well with the true $l^{*}$.}
\label{f4} 
\end{figure}

Figure \ref{f4} is an illustrative example of Lemma \ref{lem:linear} and Lemma \ref{lem:ecut}.
Using the same dataset and the two supervised classifiers as in Figure \ref{f3},
we generate $M = 10^6$ random subsets $B_m$ to estimate the empirical distribution of $\tilde{l}_{\bm{z}}$ over all $\bm{z} \in D$.
For each observation, we compute $\tilde{e}_{\bm{z}}$ and fit the linear model $\tilde{\bm{l}} = \beta\tilde{\bm{e}} + \bm{\epsilon}$ specified in Lemma \ref{lem:linear}.
The corresponding fitted values $\hat{l}_{\bm{z}}$ are obtained from this model, and the estimated cut point $\hat{l}^{*}$ is derived from the empirical estimate of $e^{*}$.
As shown in Figure~\ref{f4}, the empirical distribution of $\tilde{l}_{\bm{z}}$ closely matches with that of $\hat{l}_{\bm{z}}$,
and the estimated cut point $\hat{l}^{*}$ aligns closely with the true cut point $l^{*}$.

\subsection{The stepwise label noise cleaning algorithm (BRSLC)}

When $N$ and $M$ are sufficiently large, Lemma~\ref{lem:linear} implies that the error term $\epsilon_{\bm{z}}$ in the linear model  
\[
\tilde{l}_{\bm{z}} = \beta \tilde{e}_{\bm{z}} + \epsilon_{\bm{z}}
\]
is asymptotically negligible.  
When $N$ is smaller, however, $\epsilon_{\bm{z}}$ becomes non-negligible, making the estimated distribution $\{\tilde{l}_{\bm{z}}\}_{\bm{z}\in D}$ unstable and the threshold $e^{*}$ unreliable for separating clean and noisy observations.  

To address this, we use a stepwise procedure instead of applying a single global cut at $e^{*}$.  
At the first iteration, we apply a conservative discrimination by removing only the upper 50\% of observations among $\{\bm{z}\in D\mid \tilde{e}_{\bm{z}} \ge e^{*}\}$.  
This partial elimination provides the dataset with an initial, cautious separation that preserves most clean samples.  
In the following iterations, as the separation between clean and noisy observations becomes progressively clearer, and the influence of the error term $\epsilon_{\bm{z}}$ is reduced.  
The process is repeated until the average cross-validation error of the retained subset no longer decreases, indicating that the dataset has reached self-consistency.  
If this condition is met after the first iteration, the procedure reduces to the one-step separation, i.e., removing all $\{\bm{z}\in D\mid \tilde{e}_{\bm{z}} \ge e^{*}\}$.

The resulting algorithm, termed \emph{Label Noise Cleaning via Bernoulli Random Sampling (BRSLC)}, is summarized below.

\begin{algorithm}[H]
\caption{\emph{Label Noise Cleaning via Bernoulli Random Sampling} (BRSLC)}
\label{alg:brslc}
\begin{algorithmic}[1]
\Require Noisy dataset $D=\{\bm{z}_i=(\bm{x}_i,y_i)\}_{i=1}^{N}$; classifier $h$; Bernoulli rate $q\in(0,1)$; number of subsets $M$
\Ensure Cleaned dataset $D_{\mathrm{cleaned}}$

\State $t = 1$, \ $D^{(t)} = D$, \ $e_{cv}(h, D^{*(0)}) = +\infty$
\While{true}
  \State For each $\bm{z}_i \in D^{(t)}$, initialize $s_i = 0$, $n_i = 0$
  \For{$m = 1,\dots,M$}
    \State Sample a Bernoulli random subset $B_m \subseteq D^{(t)}$ with inclusion rate $q$
    \State Train $h_{B_m}$ and compute the cross-validation error $e_{cv}(h, B_m)$
    \ForAll{$\bm{z}_i \in B_m$}
      \State $s_i = s_i + e_{cv}(h, B_m)$, \quad $n_i = n_i + 1$
    \EndFor
  \EndFor
  \State For all $n_i > 0$, set $\tilde{e}_{\bm{z}_i} = s_i / n_i$
  \State Compute the cut point $e_t^{*}$ from $\{\tilde{e}_{\bm{z}_i} : \bm{z}_i \in D^{(t)}\}$ using Lemma~\ref{lem:ecut}
  \State $\hat{D}^{(t)}_1 = \{\bm{z}_i \in D^{(t)} : \tilde{e}_{\bm{z}_i} < e_t^{*}\}$
  \State Train $h_{\hat{D}^{(t)}_1}$ and compute $e_{cv}(h, \hat{D}^{(t)}_1)$
  \If{$e_{cv}(h, \hat{D}^{(t)}_1) \ge e_{cv}(h, \hat{D}^{(t-1)}_1)$}
    \State \Return $D_{\mathrm{cleaned}} = \hat{D}^{(t-1)}_1$
  \EndIf
  \State $\hat{D}^{(t)}_2 = D^{(t)} \setminus \hat{D}^{(t)}_1$
  \State $\tau_t = \mathrm{median}\{\tilde{e}_{\bm{z}_i} : \bm{z}_i \in \hat{D}^{(t)}_2\}$
  \State $S^{(t)} = \{\bm{z}_i \in \hat{D}^{(t)}_2 : \tilde{e}_{\bm{z}_i} > \tau_t\}$
  \State $D^{(t+1)} = D^{(t)} \setminus S^{(t)}$, \quad $t = t + 1$
\EndWhile
\end{algorithmic}
\end{algorithm}

\begin{figure}[t]
    \centering

    \begin{minipage}[t]{0.48\textwidth}
        \centering
        \includegraphics[width=\linewidth,height=5.4cm]{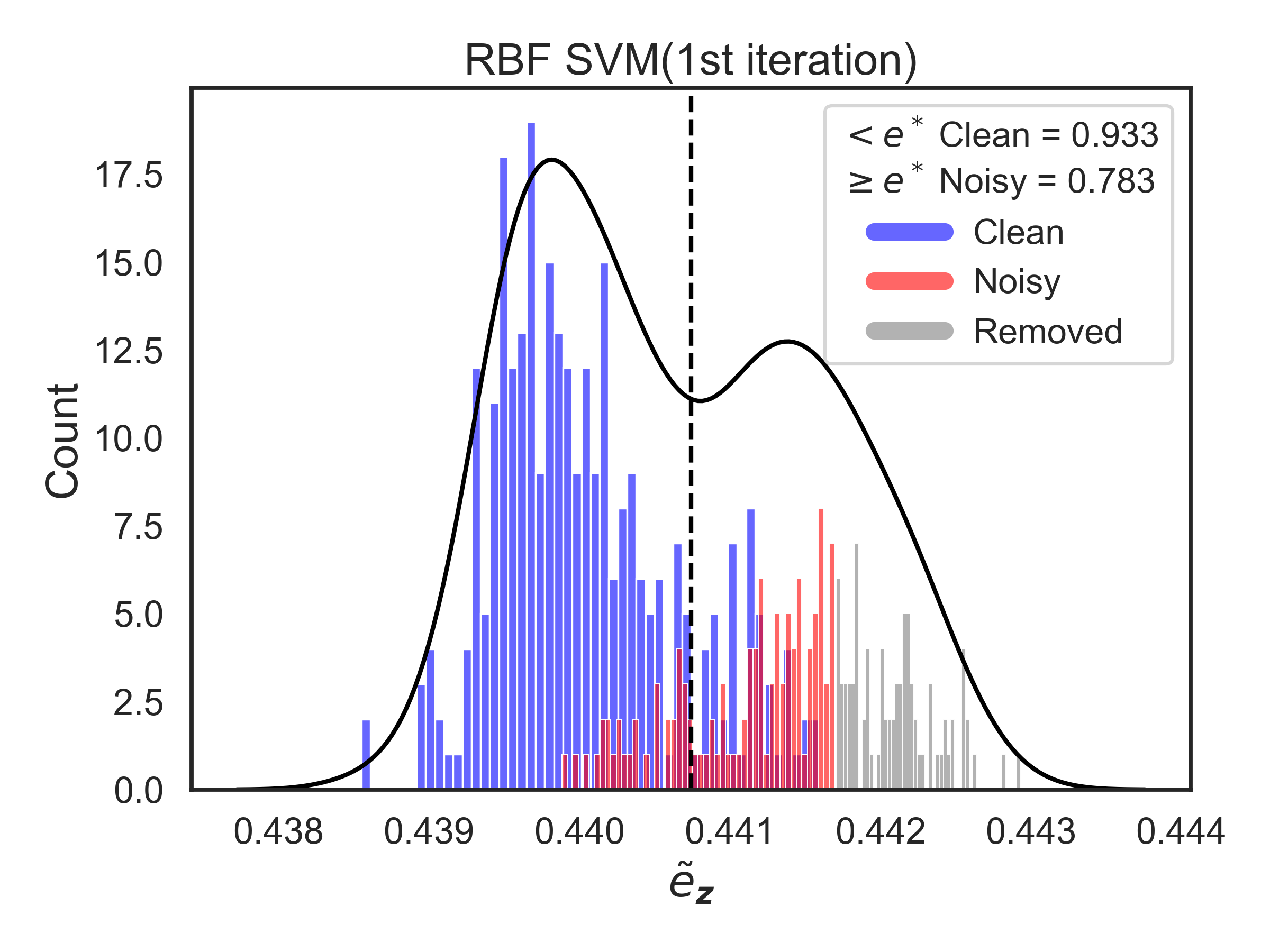}
    \end{minipage}\hspace{0.01\textwidth}
    \begin{minipage}[t]{0.48\textwidth}
        \centering
        \includegraphics[width=\linewidth,height=5.4cm]{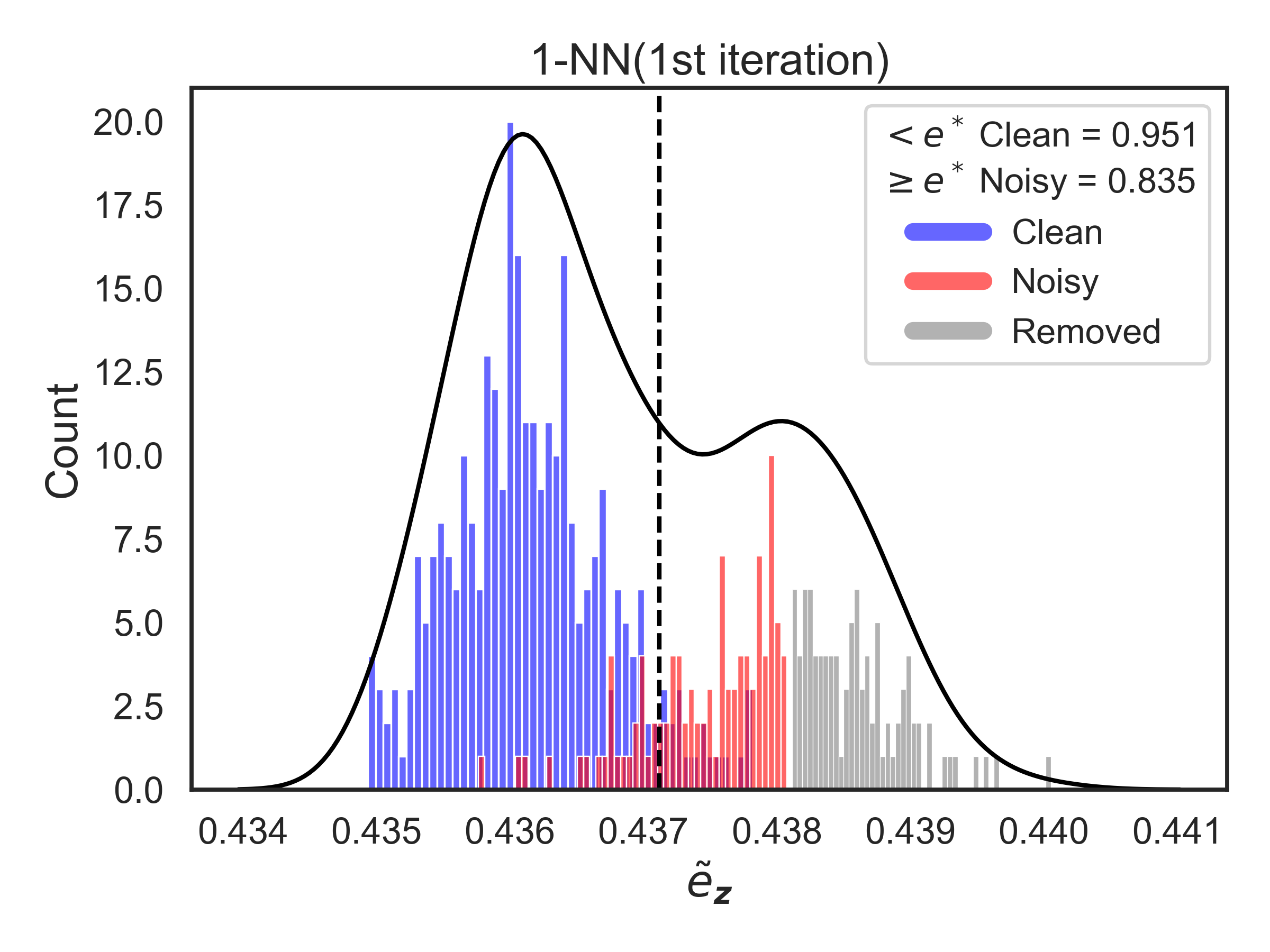}
    \end{minipage}

    \vspace{0.6em}

    \begin{minipage}[t]{0.48\textwidth}
        \centering
        \includegraphics[width=\linewidth,height=5.4cm]{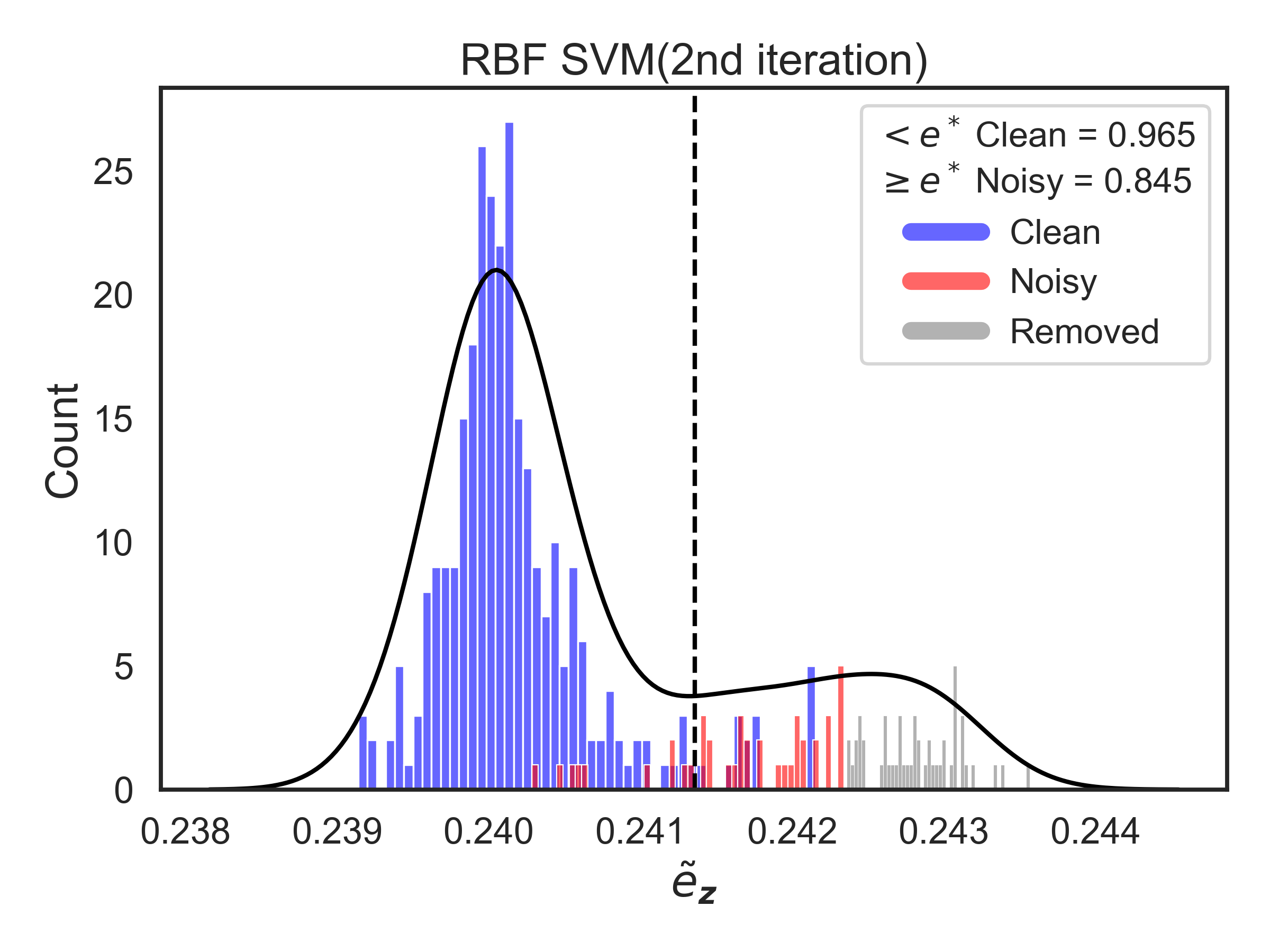}
    \end{minipage}\hspace{0.01\textwidth}
    \begin{minipage}[t]{0.48\textwidth}
        \centering
        \includegraphics[width=\linewidth,height=5.4cm]{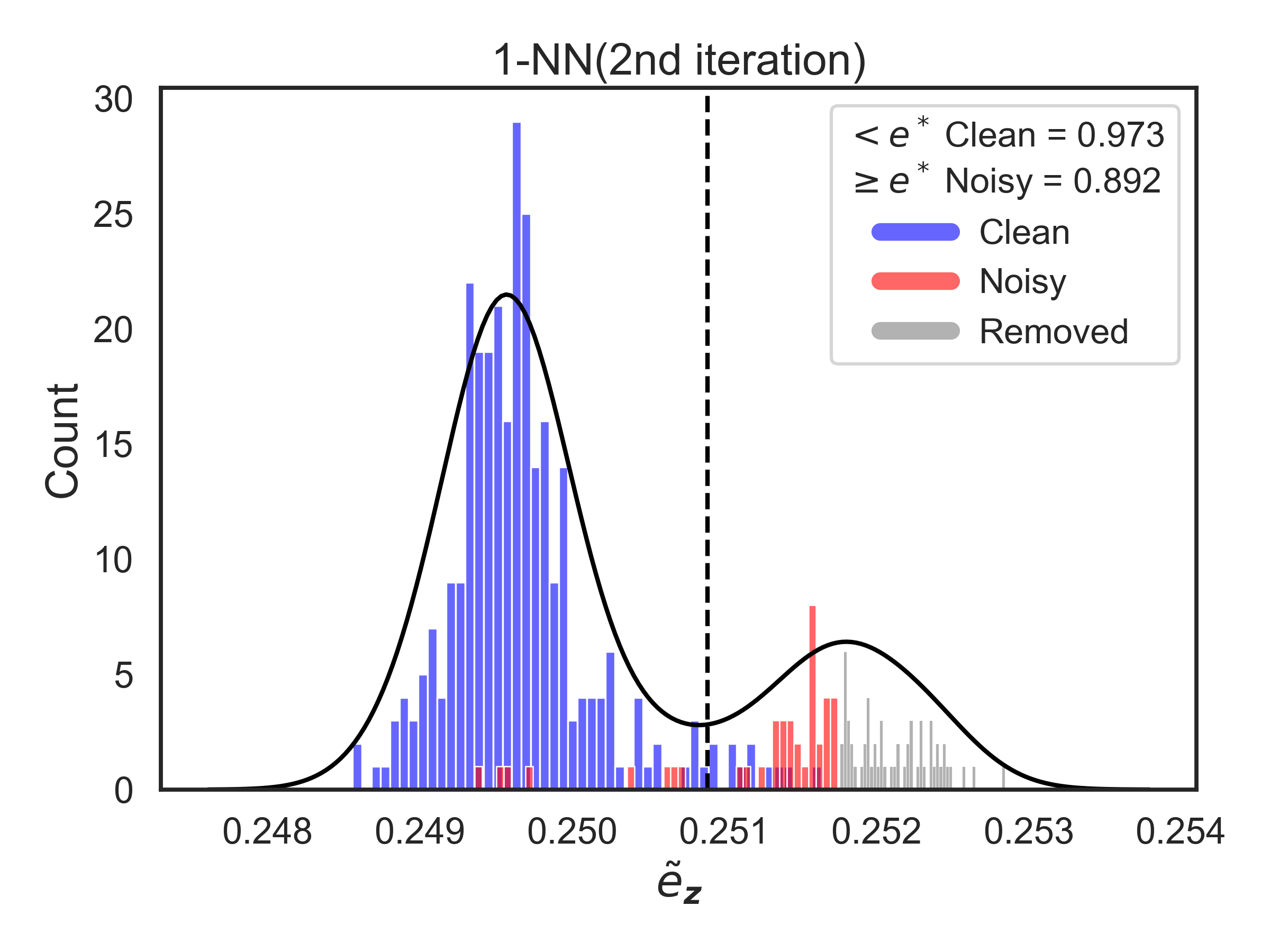}
    \end{minipage}

    \vspace{0.6em}

    \begin{minipage}[t]{0.48\textwidth}
        \centering
        \includegraphics[width=\linewidth,height=5.4cm]{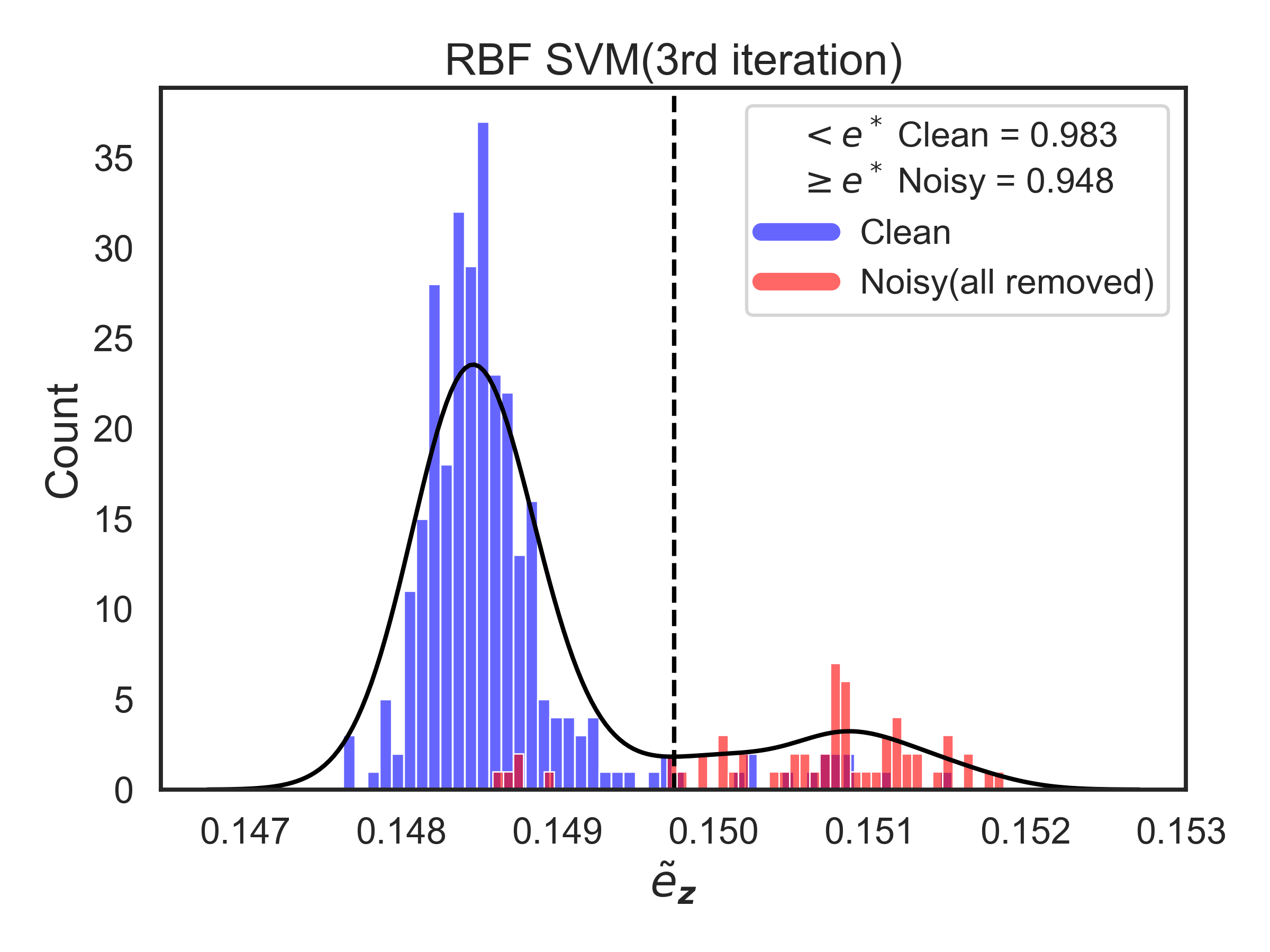}
    \end{minipage}\hspace{0.01\textwidth}
    \begin{minipage}[t]{0.48\textwidth}
        \centering
        \includegraphics[width=\linewidth,height=5.4cm]{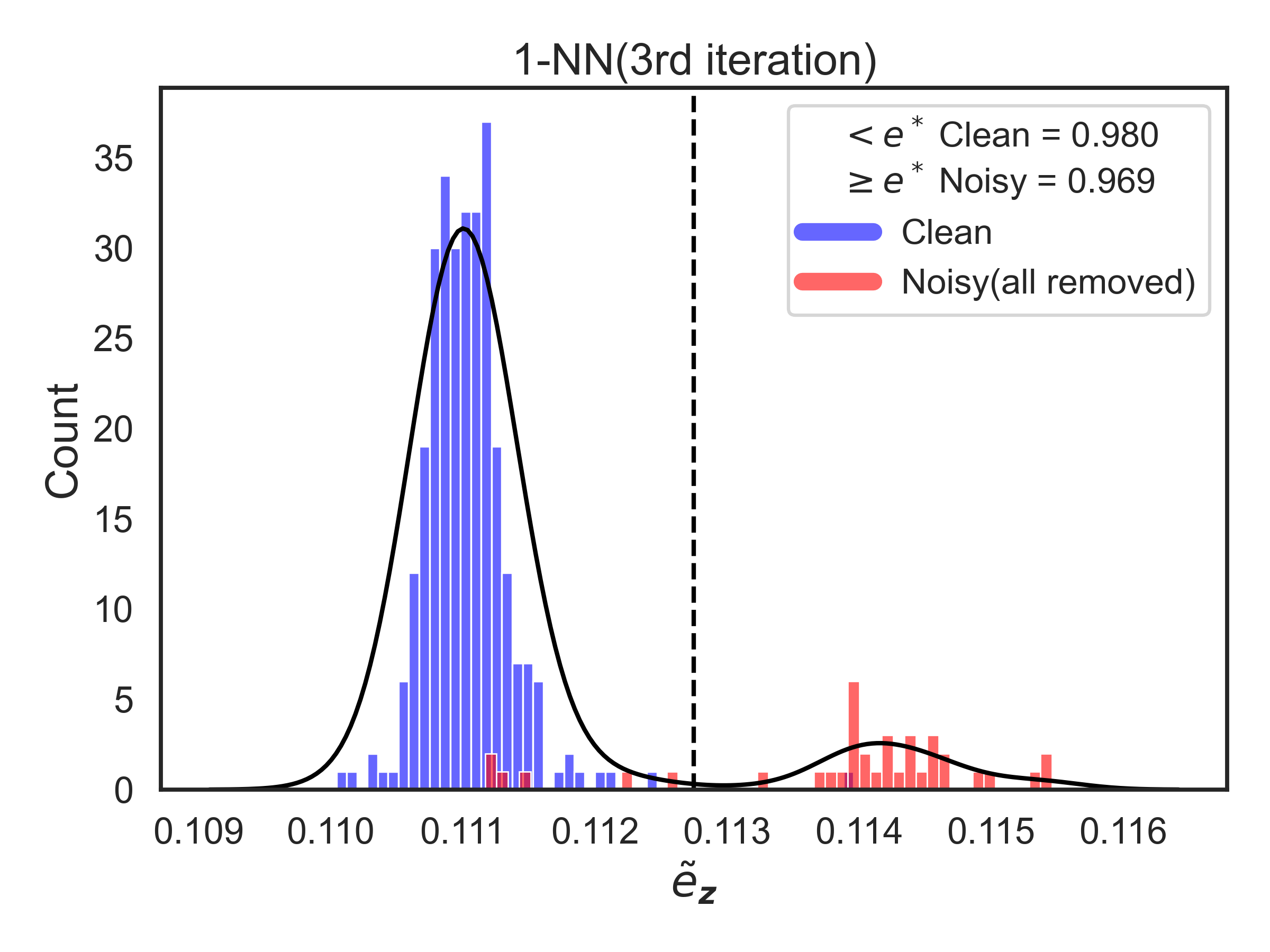}
    \end{minipage}

    \caption{Three-iteration BRSLC performance of RBF SVM and 1-NN. The legend $< e^{*}$ denotes $\frac{\#\{\bm{z}\in D_1\mid \tilde{e}_{\bm{z}} < e^{*}\}}{\#\{\bm{z}\in D\mid \tilde{e}_{\bm{z}} < e^{*}\}}$, $\ge e^{*}$ denotes $\frac{\#\{\bm{z}\in D_2\mid \tilde{e}_{\bm{z}} \ge e^{*}\}}{\#\{\bm{z}\in D\mid \tilde{e}_{\bm{z}} \ge e^{*}\}}$.}
    \label{f5}
\end{figure}

Figures \ref{f5} is an illustrative example of BRSLC.
Using the same data distribution and the two supervised classifiers as in Figure \ref{f3}, this time we reduce the size of dataset $N$ from 1000 to 500, and improve the original label noise level to 40\%. BRSLC is repeated for 3 iterations on this dataset for both classifiers. As shown in the figures, during the first iteration, the proportion of noisy observations among those satisfying $\tilde{e}_{\bm{z}} \ge e^{*}$,  
i.e., $\frac{\#\{\bm{z}\in D_2\mid \tilde{e}_{\bm{z}} \ge e^{*}\}}{\#\{\bm{z}\in D\mid \tilde{e}_{\bm{z}} \ge e^{*}\}}$, is 78.3\% and 83.5\% for the two classifiers, respectively.  
As the iterations proceed, the proportion of misidentified clean samples decreases, and by the third iteration—where the stopping criterion is met, $\frac{\#\{\bm{z}\in D_2\mid \tilde{e}_{\bm{z}} \ge e^{*}\}}{\#\{\bm{z}\in D\mid \tilde{e}_{\bm{z}} \ge e^{*}\}}$ is improved to 94.8\% and 96.9\%. Similarly, the proportion of clean observations among those with $\tilde{e}_{\bm{z}} < e^{*}$, i.e., $\frac{\#\{\bm{z}\in D_1\mid \tilde{e}_{\bm{z}} < e^{*}\}}{\#\{\bm{z}\in D\mid \tilde{e}_{\bm{z}} < e^{*}\}}$, reaches 98.3\% and 98.0\%.  
These results demonstrate the necessity of the stepwise framework for improving label noise identification accuracy.

In Section~\ref{sec4}, we evaluate the performance of BRSLC on both simulated and real datasets, comparing it against existing label cleaning approaches in terms of clean data retention, residual noise reduction, and improvement in downstream classification accuracy.

\section{Experiments}\label{sec4}
\subsection{Experiments on simulated data with artificial label noise}
We first evaluate our method on four simulated data settings (Settings~1–4) described below.
\par\vspace{4\baselineskip}
\textbf{Setting 1:}  
\[
X^{-} \sim \mathcal{N}(\bm{0}, \Sigma), 
\quad 
X^{+} \sim \mathcal{N}(\bm{1}, \Sigma),
\]
with
\[
\Sigma =
\begin{pmatrix}
2 & -1 & 0 \\
-1 & 2 & -1 \\
0 & -1 & 2
\end{pmatrix},
\qquad 
N^{-} : N^{+} = 1 : 1.
\]

\textbf{Setting 2:}  
\[
X^{-}, \; X^{+} \sim t_{\nu}(\mu, \Sigma),
\]
with degrees of freedom $\nu = 4$, the same covariance matrix $\Sigma$ as in Setting~1,  
\[
\mu^{-} = \bm{0}, 
\qquad 
\mu^{+} = \bm{1}, 
\qquad 
N^{-} : N^{+} = 1 : 1.
\]

\textbf{Setting 3:}  
The negative class $X^{-}$ is drawn from a uniform distribution on a circle centered at $(0,0)^T$ or $(2,2)^T$, each with probability $1/2$.  
The positive class $X^{+}$ is drawn from a uniform distribution on a circle centered at $(1,1)^T$.  
The sample size ratio is
\[
N^{-} : N^{+} = 1 : 2.
\]

\textbf{Setting 4:}  
\[
\begin{aligned}
& X \in \mathbb{R}^{10}, \qquad x_3, \dots, x_{10} \sim \mathcal{N}(0,1), \\[0.5ex]
& x_1 \sim 
\begin{cases}
\mathcal{N}(1, 0.5), & y = 1, \\
\mathcal{N}(-1, 1), & y = 0,
\end{cases}
\qquad
x_2 \sim 
\begin{cases}
\mathcal{N}(-1, 0.5), & y = 1, \\
\mathcal{N}(1, 1), & y = 0,
\end{cases} \\[1ex]
&\quad N^{-} : N^{+} = 3 : 7.
\end{aligned}
\]

For each setting, we randomly generate 50 independent training datasets, each consisting of 1000 observations.  
Symmetric label noise is introduced by randomly flipping a specified proportion of labels, with noise levels set at 10\%, 20\%, 30\%, and 40\%.  

Table~\ref{t1} shows the performance of the proposed cleaning method across four artificial data settings and four levels of injected label noise.
The reported values are averaged over three types of classifiers (RBF SVM, CART, and $1$-NN).
For each setting, we compare the original noise level $\ell(D)$ with the residual label noise level after applying BRSLC ($\ell_{\text{cleaned}}$). In addition, we report the proportion of noisy observations successfully removed ($r_{\text{noise}}$), and the proportion of clean observations retained in the cleaned dataset ($r_{\text{clean}}$).

The results that BRSLC substantially reduces the label noise level $\ell(D)$ across all four settings. 
When the original noise level is between $10\%$ and $30\%$, $\ell_{\text{cleaned}}$ is typically reduced to between about $0.5\%$ and $7.4\%$, corresponding to removing roughly $85\%$–$95\%$ of the noisy observations, while retaining at least $86\%$–$97\%$ of the clean observations. 
Settings~1 and~4 are the most favorable, with $\ell_{\text{cleaned}}$ generally below $3\%$ up to $\ell(D)=30\%$ and $r_{\text{clean}}$ above $90\%$.

At the highest label noise level $\ell(D)=40\%$, the residual label noise level remains low in Settings~1 and~4 
($\ell_{\text{cleaned}}=3.40\%$ and $7.12\%$), and BRSLC continues to remove a large portion of noisy observationss 
($r_{\text{noise}}=91.5\%$ and $82.2\%$).  
In Settings~2 and~3, although the residual label noise level increases ($\ell_{\text{cleaned}}=10.43\%$ and $16.20\%$), 
the method still removes a substantial fraction of noisy observations ($r_{\text{noise}}=73.9\%$ and $59.5\%$) 
while retaining the majority of clean observations.  
Overall, even under this challenging 40\% noise scenario, the procedure preserves a large clean subset, 
with $r_{\text{clean}}$ remaining above $77\%$ across all four settings.

\begin{table}[t]
\centering
\small
\setlength{\tabcolsep}{8pt}
\renewcommand{\arraystretch}{1.2}
\begin{tabular}{ccccc}
\toprule
\textbf{Setting} & $\boldsymbol{\ell(D)}$ (\%) & $\boldsymbol{\ell_{\text{cleaned}}}$ (\%) & 
$\boldsymbol{r_{\text{noise}}}$ (\%) & $\boldsymbol{r_{\text{clean}}}$ (\%) \\
\midrule
1 & 10 & 0.54 & 94.6 & 96.7 \\
  & 20 & 1.18 & 94.1 & 93.2 \\
  & 30 & 2.62 & 91.3 & 92.0 \\
  & 40 & 3.40 & 91.5 & 88.6 \\
\midrule
2 & 10 & 1.06 & 89.4 & 95.7 \\
  & 20 & 2.20 & 89.0 & 92.6 \\
  & 30 & 4.17 & 86.1 & 89.5 \\
  & 40 & 10.43 & 73.9 & 80.1 \\
\midrule
3 & 10 & 1.55 & 84.5 & 86.9 \\
  & 20 & 3.86 & 80.7 & 85.6 \\
  & 30 & 7.39 & 75.4 & 84.0 \\
  & 40 & 16.20 & 59.5 & 77.7 \\
\midrule
4 & 10 & 1.03 & 89.7 & 95.6 \\
  & 20 & 1.59 & 92.1 & 94.6 \\
  & 30 & 2.76 & 90.8 & 91.0 \\
  & 40 & 7.12 & 82.2 & 84.3 \\
\bottomrule
\end{tabular}
\caption{Summary of label noise cleaning performance across four simulated data settings. 
All values are expressed as percentages. 
$\ell(D)$: original label noise level; 
$\ell_{\text{cleaned}}$: residual label noise level after cleaning; 
$r_{\text{noise}} = 100 \times [1 - \ell_{\text{cleaned}} / \ell(D)]$: proportion of label noise removed; 
$r_{\text{clean}} = 100 \times (n_{\text{cleaned}} / n_{\text{clean}})$: proportion of clean observations retained after cleaning.}
\label{t1}
\end{table}

To evaluate classification performance before and after label cleaning, we generate a clean test set of size 500 from the corresponding clean data distribution for each simulated training dataset.  

\begin{figure}[t]
  \centering
  \begin{tabular}{@{}c@{\hspace{0.06\textwidth}}c@{}}
    \includegraphics[width=0.46\textwidth]{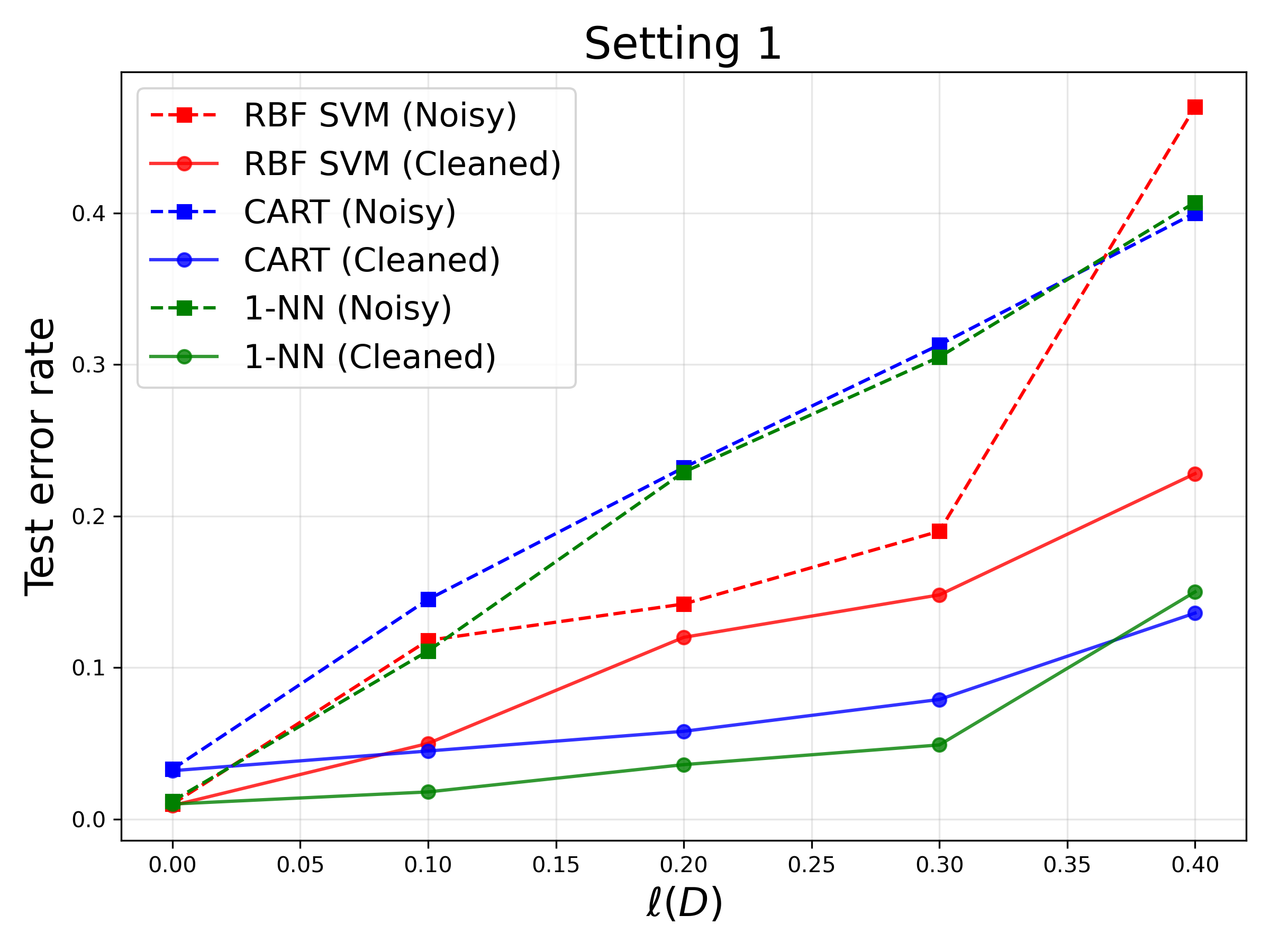} &
    \includegraphics[width=0.46\textwidth]{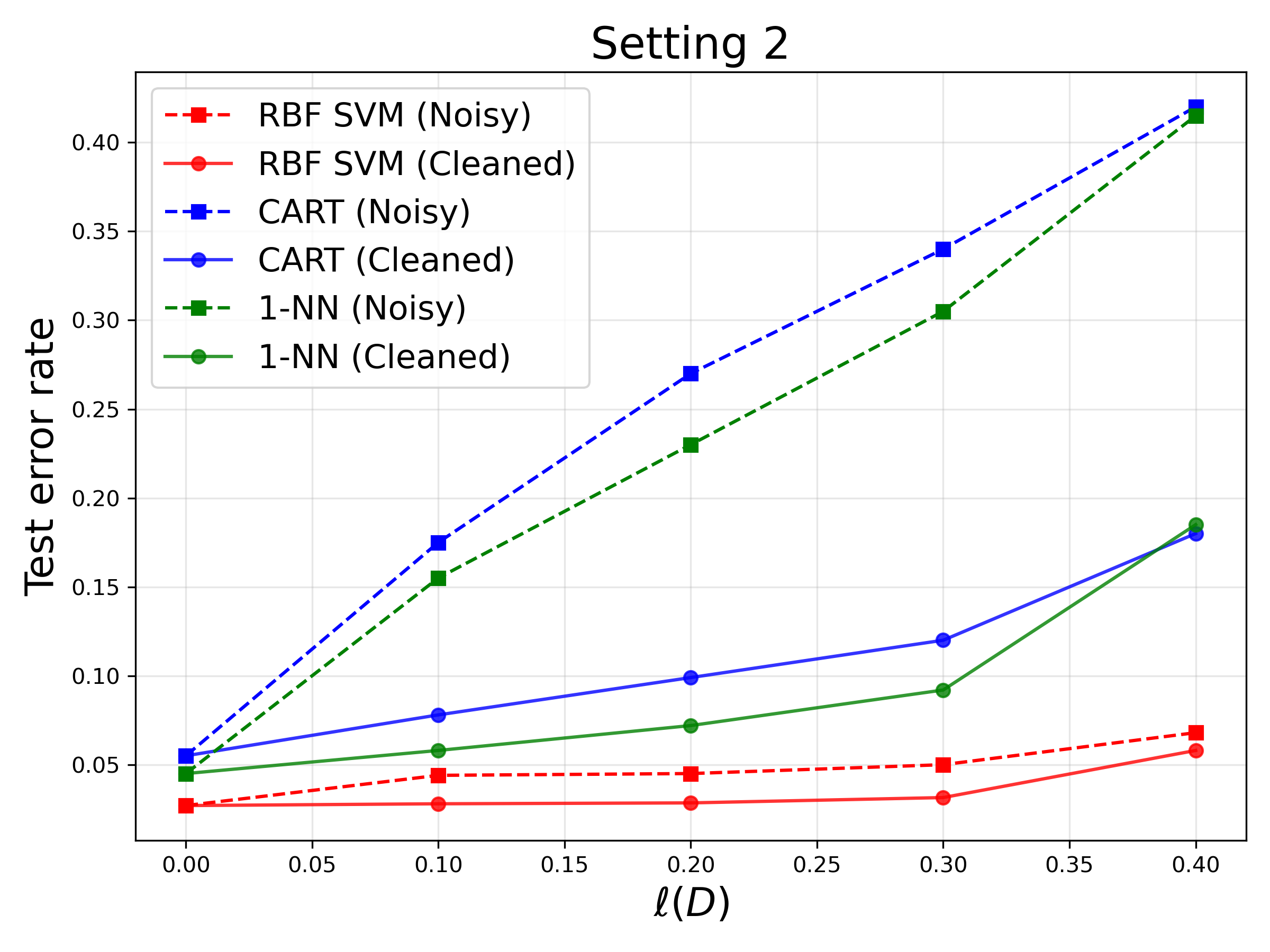} \\
    [0.8em]
    \includegraphics[width=0.46\textwidth]{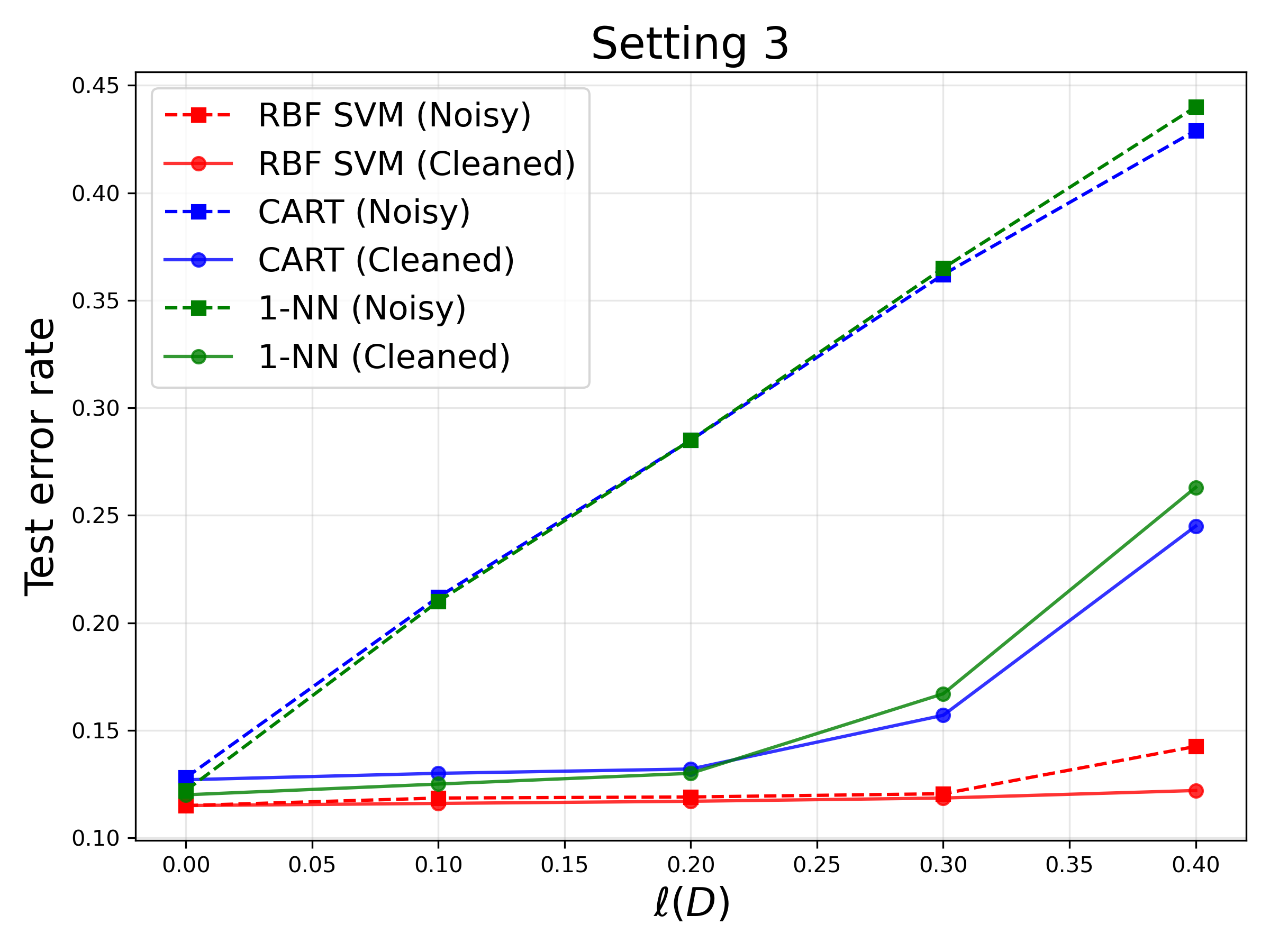} &
    \includegraphics[width=0.46\textwidth]{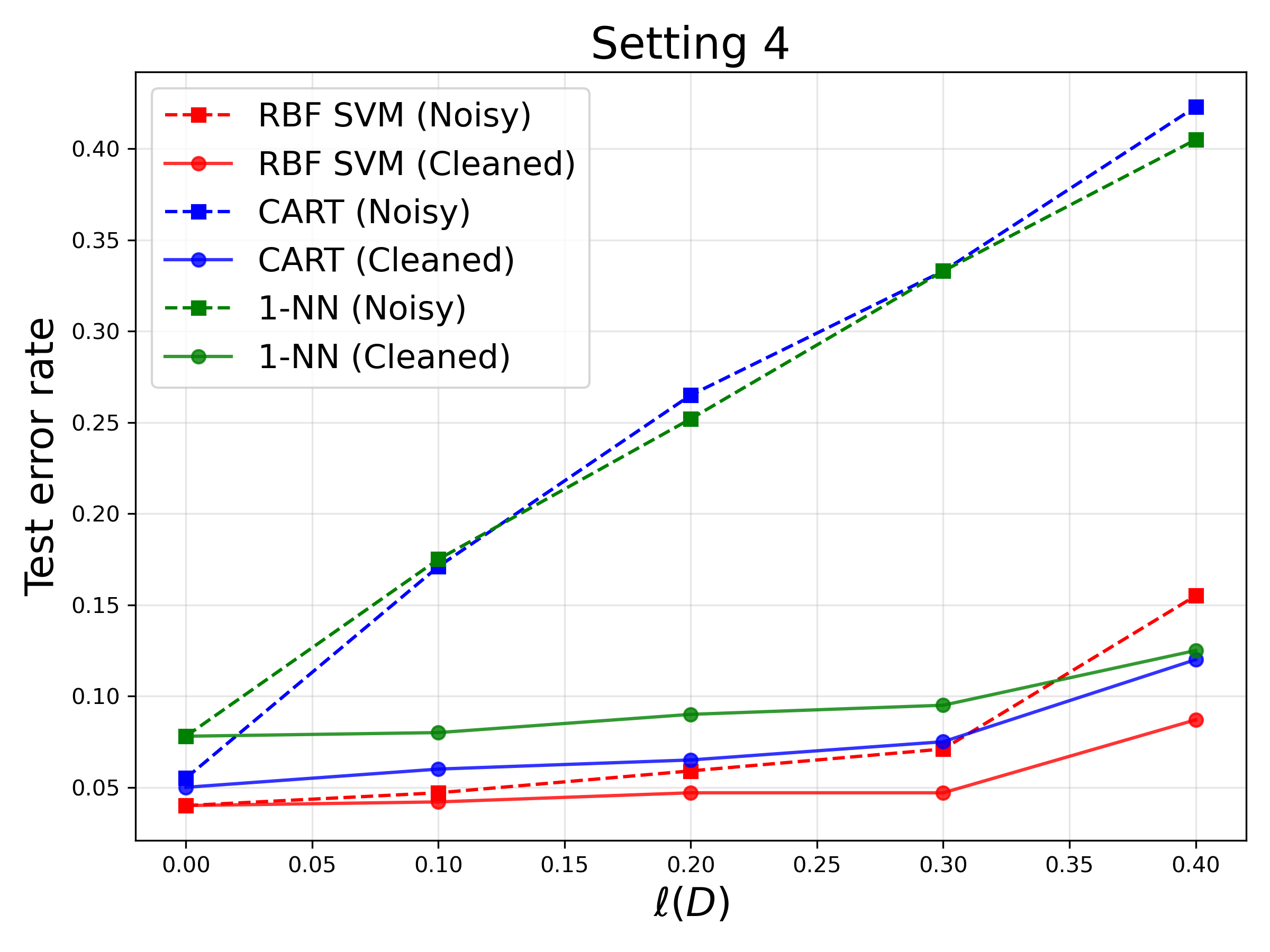}
  \end{tabular}

  \caption{Test error rates of noisy and cleaned datasets across the four simulated data settings.  
Each panel corresponds to one setting, showing results for three classifiers: RBF SVM, CART, and 1-NN.  
Solid lines represent performance on after BRSLC cleaning, while dashed lines represent performance of original noisy data.}
\label{f6}
\end{figure}

Figure~\ref{f6} shows the average test error rates of three supervised classifiers—RBF SVM, CART, and 1-NN—across the four simulated settings, comparing performance on noisy versus cleaned training data.  
Across noise levels from $10\%$ to $40\%$, the maximum reduction in average test error ranges from $1.9\%$ to $14.2\%$ for RBF SVM, from $18.4\%$ to $30.3\%$ for CART, and from $19.8\%$ to $27.9\%$ for 1-NN.

Detailed results corresponding to Table~\ref{t1} and Figure~\ref{f6} are provided in the Supplementary Materials (Table~S1 and Figures~S1–S3), which include the experiment results across all simulated settings, classifiers, and noise levels.

\subsection{Experiments on real-world data with artificial label noise}
We apply BRSLC on four real-world datasets. The Wisconsin Diagnostic Breast Cancer (WDBC) dataset contains 569 observations\citep{street1993nuclear}, each represented by a 30-dimensional feature vector and labeled as either benign or malignant. The Mushroom dataset\citep{mushroom_73} comprises 8,124 observations with 23-dimensional feature vectors and labels indicating whether the mushroom is edible or poisonous. The UCI Letter Recognition dataset\citep{letter_recognition_59} includes 20,000 observations, each with a 16-dimensional feature vector and a label from A to Z. The MNIST digit dataset\citep{borji2008robust} consists of 60,000 observations, each represented by a 16-dimensional feature vector and labeled from 0 to 9.

\begin{figure}[t]
  \centering
  \begin{tabular}{@{}c@{}}
    \includegraphics[width=0.7\textwidth]{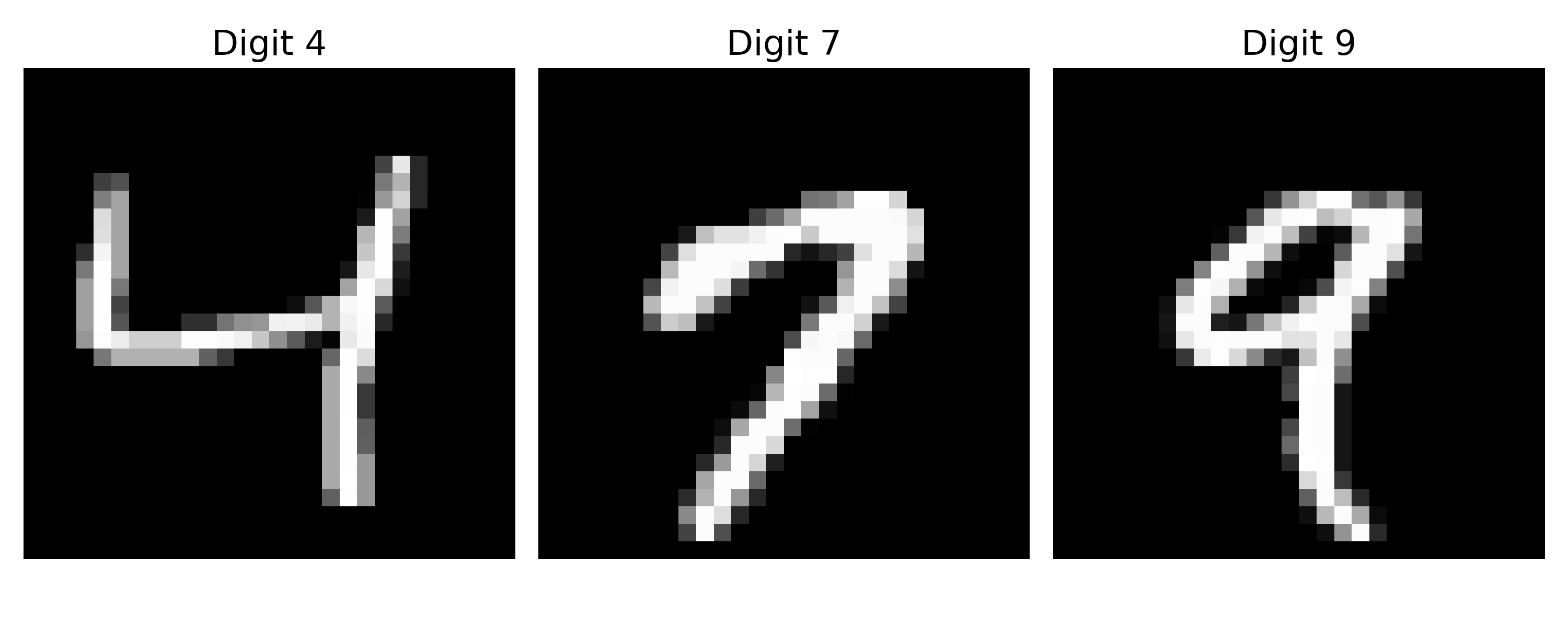} \\
    [\baselineskip]
    \includegraphics[width=0.7\textwidth]{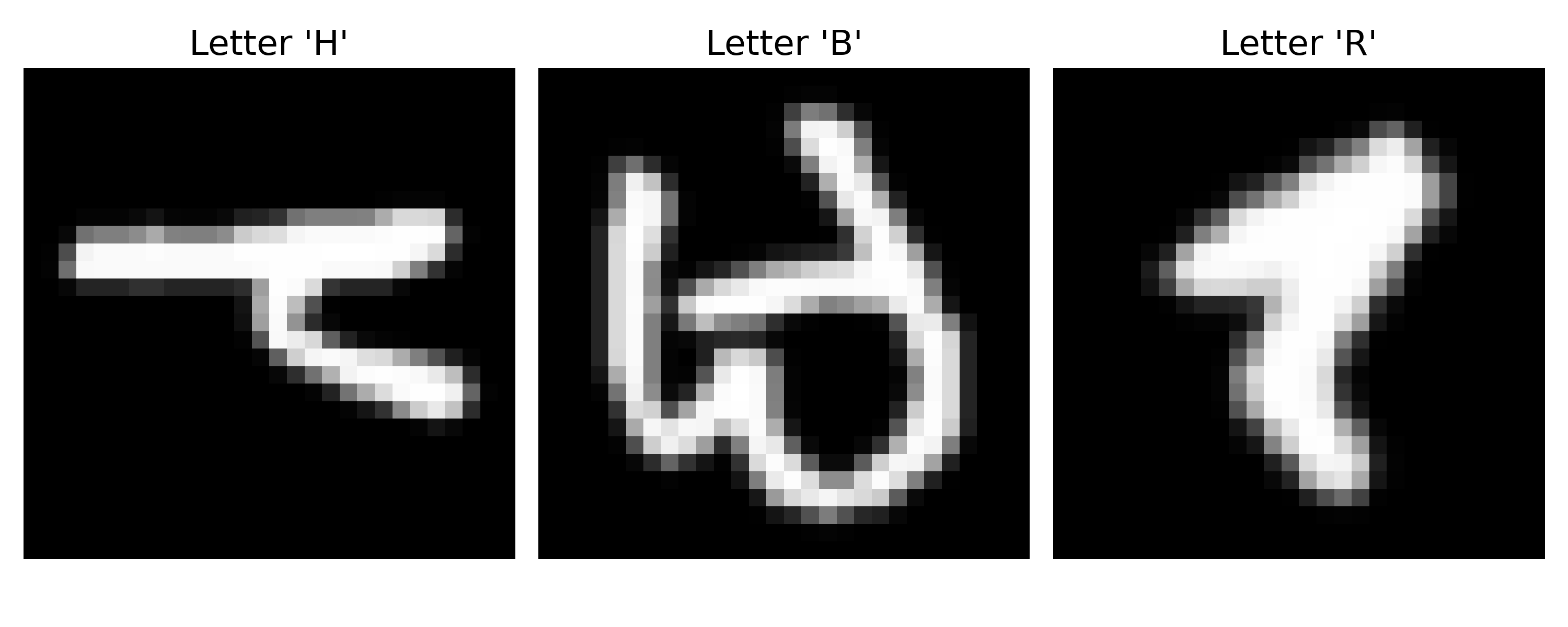}
  \end{tabular}
  \caption{Examples of letters B, H, R in dataset UCI letter
  recognition, and digits 4, 7, 9 in dataset MNIST digit recognition.}
  \label{f7}
\end{figure}

For the UCI Letter Recognition dataset, the letters B, H, and R are among the most difficult to classify. Similarly, in the MNIST dataset, digits 4, 7, and 9 are often confused with one another (Figure~\ref{f7} provides an example). Therefore, in our study, we focus on these subsets: the three letters (2,258 observations) from the UCI dataset and the three digits (18,056 observations) from the MNIST dataset.

We perform eight classification experiments using data sampled from the four real-world datasets. In each experiment, we construct a binary classification task by selecting a subset of relevant classes from the original dataset and randomly sampling training and test sets with specified class proportions.

For example, in the first experiment, we sample 400 observations from the WDBC dataset for training and 100 for testing, with labels “Malignant” as class 0 and “Benign” as class 1. In the third experiment, based on the UCI Letter Recognition dataset, we sample 500 instances of the letter B, 250 of H, and 250 of R to form a training set of 1,000 observations. Here, letter B is designated as class 0 and letters H and R as class 1. A corresponding test set of 200 observations is similarly sampled from the remaining data. Table~\ref{t2} gives a summary of all experiment settings.
\begin{table}[t]
\centering
\renewcommand{\arraystretch}{1.2}
\begin{tabular}{l l l c c}
\toprule
\textbf{Experiment} & \textbf{Label 0} & \textbf{Label 1} & \boldmath$n_{\text{train}}$ & \boldmath$n_{\text{test}}$ \\
\midrule
WDBC        & Malignant & Benign       & 400  & 100  \\
Mushroom    & Edible    & Poisonous    & 1000 & 500  \\
Letter-B    & B         & H and R      & 1000 & 200  \\
Letter-H    & H         & B and R      & 1000 & 200  \\
Letter-R    & R         & B and H      & 1000 & 200  \\
MNIST-4     & 4         & 7 and 9      & 2000 & 1000 \\
MNIST-7     & 7         & 4 and 9      & 2000 & 1000 \\
MNIST-9     & 9         & 4 and 7      & 2000 & 1000 \\
\bottomrule
\end{tabular}
\caption{Summary of the eight experiments in our study. Each row defines a binary classification task derived from an underlying dataset.  
For the Letter and MNIST subsets, the total sample sizes for Label~0 and Label~1 follow a 50:50 split.  
Within Label~1, the two constituent classes are balanced in a 50:50 ratio as well.}
\label{t2}
\end{table}

For each experiment, we introduce label noise at levels of 10\%, 20\%, 30\%, and 40\%, and independently repeat each experiment at each noise level 50 times. We evaluate classification performance using two models: Support Vector Machines (SVM) with a linear kernel (Linear SVM) and with a radial basis function kernel (RBF SVM). Our method is compared against the approach proposed by Ekambaram et al. (2016), which introduces a label noise purification technique known as Active Label Noise Removal (ALNR).

ALNR follows a stepwise framework based on the observation that noisy samples are more likely to appear as support vectors. In each iteration, the dataset is partitioned into support vectors and non-support vectors. A new SVM is trained using only the non-support vectors, which is then used to classify the support vectors. The support vectors with the highest misclassification probabilities are removed, and this process is repeated until no misclassified observations remain.

Table~\ref{t3} gives the average performance of ALNR and BRSLC under four label noise levels, aggregated across all datasets and evaluated with both Linear SVM and RBF SVM.
Overall, BRSLC shows clear advantages: it achieves a lower averaged residual label noise level 
($\ell_{\text{cleaned}}$) than ALNR in all 8 out of 8 comparisons, and it retains a higher proportion 
of clean observations ($r_{\text{clean}}$) in all 8 cases.  
At the highest noise level, $\ell(D)=40\%$, the advantage is the most evident: 
BRSLC reduces $\ell_{\text{cleaned}}$ by 4.8\% for Linear SVM and by 5.0\% for RBF SVM, 
and increases $r_{\text{clean}}$ by 5.7\% and 7.0\%, respectively.

\begin{table}[t]
\centering
\small
\setlength{\tabcolsep}{8pt}
\renewcommand{\arraystretch}{1.2}
\begin{tabular}{ccccc}
\toprule
$\boldsymbol{\ell(D)}$ (\%) & \textbf{Method} & \textbf{Classifier} &
$\boldsymbol{\ell_{\text{cleaned}}}$ (\%) & $\boldsymbol{r_{\text{clean}}}$ (\%) \\
\midrule
10 & ALNR  & Linear SVM & 1.1  & 89.4 \\
   &       & RBF SVM    & 1.0  & 86.0 \\
   & BRSLC & Linear SVM & \textbf{1.0} & \textbf{90.1} \\
   &       & RBF SVM    & \textbf{0.5} & \textbf{89.6} \\
\midrule
20 & ALNR  & Linear SVM & 2.9  & 85.9 \\
   &       & RBF SVM    & 2.4  & 83.0 \\
   & BRSLC & Linear SVM & \textbf{2.5} & \textbf{87.3} \\
   &       & RBF SVM    & \textbf{1.5} & \textbf{87.5} \\
\midrule
30 & ALNR  & Linear SVM & 7.1  & 80.2 \\
   &       & RBF SVM    & 5.4  & 78.0 \\
   & BRSLC & Linear SVM & \textbf{5.5} & \textbf{81.0} \\
   &       & RBF SVM    & \textbf{3.2} & \textbf{84.0} \\
\midrule
40 & ALNR  & Linear SVM & 18.4 & 74.0 \\
   &       & RBF SVM    & 16.3 & 71.0 \\
   & BRSLC & Linear SVM & \textbf{13.6} & \textbf{79.7} \\
   &       & RBF SVM    & \textbf{11.3} & \textbf{78.0} \\
\bottomrule
\end{tabular}
\caption{Average results of ALNR and BRSLC across eight datasets under different noise levels $\ell(D)$. 
$\ell_{\text{cleaned}}$: residual label noise level after cleaning; 
$r_{\text{clean}}$: proportion of clean observations retained after cleaning, expressed as a percentage. 
All values are averaged over datasets and classifiers.}
\label{t3}
\end{table}

\begin{figure}[t] \centering \begin{tabular}{@{}c@{\hspace{0.06\textwidth}}c@{}} \includegraphics[width=0.48\textwidth]{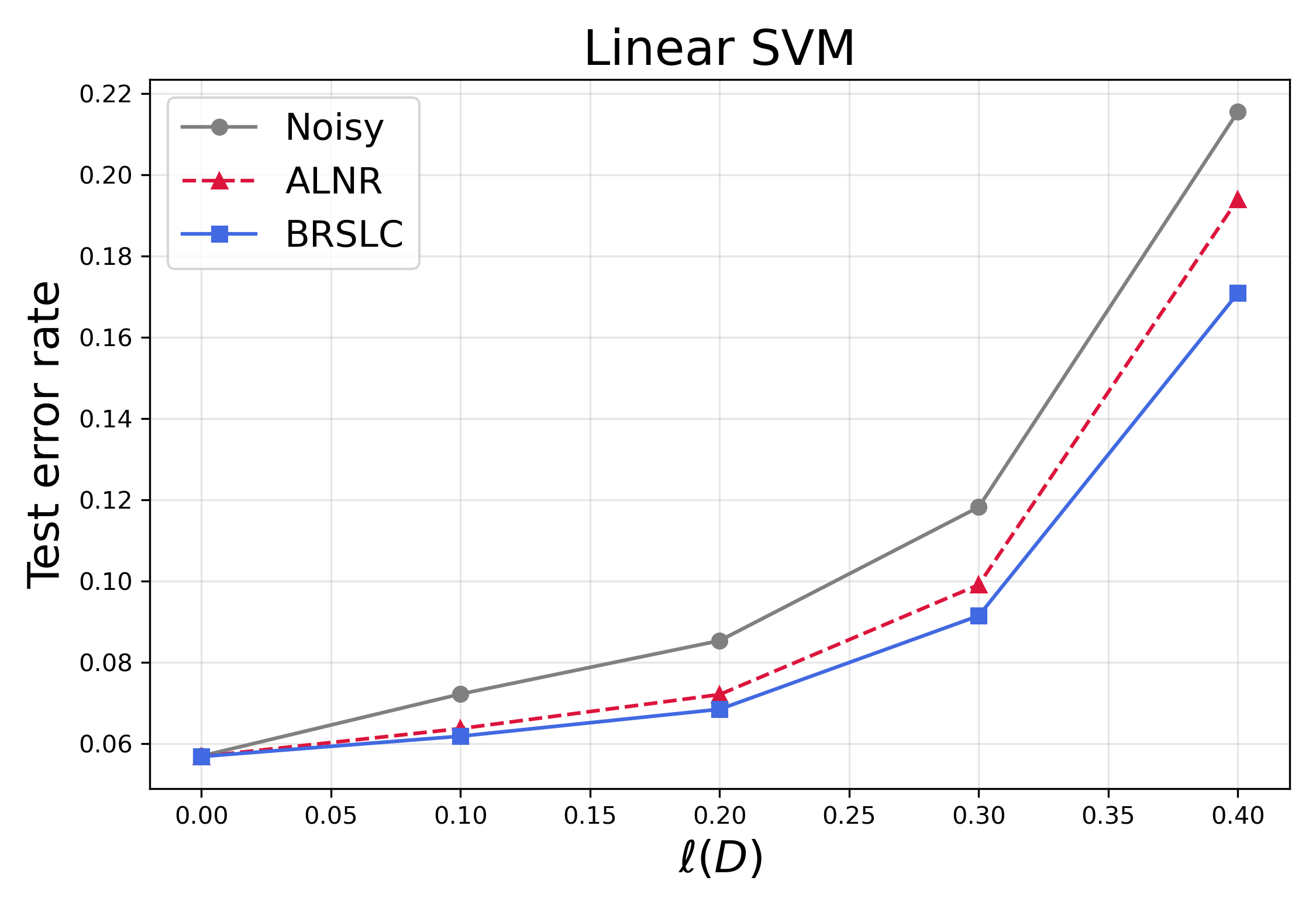} & \includegraphics[width=0.48\textwidth]{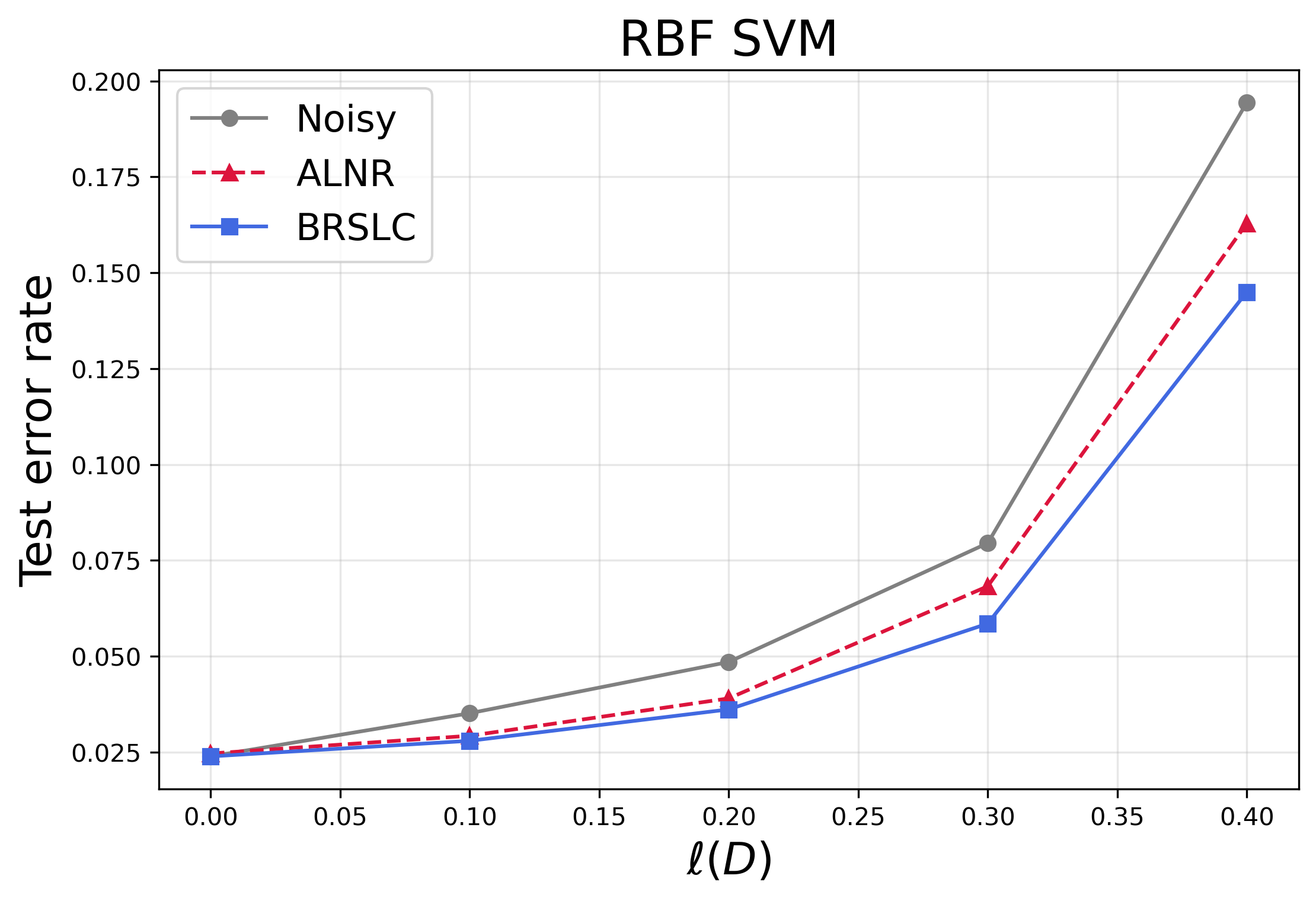} \\ \end{tabular} \caption{Average test error rates of Linear SVM trained on noisy data, and data cleaned by ALNR and BRSLC.} \label{f8} 
\end{figure}

Figure~\ref{f8} shows the averaged test error rates across eight datasets for the original noisy data and the data cleaned by ALNR and BRSLC, using Linear SVM (left) and RBF SVM (right) as classifiers.  
Both cleaning methods substantially reduce test error compared to training on the noisy data.  
At low noise levels ($\ell(D) = 10\%$ and $20\%$), the performance of ALNR and BRSLC is nearly identical, with differences below 0.5\%.  
However, at higher noise levels, BRSLC consistently outperforms ALNR.  
For Linear SVM, when $\ell(D) = 30\%$ and $40\%$, BRSLC achieves 0.76\% and 2.31\% lower test error than ALNR, respectively.  
For RBF SVM, the corresponding improvements are 0.98\% and 1.79\%.  

The results in Table~\ref{t3} and Figure~\ref{f8} show that BRSLC performs better than ALNR in cleaning label noise, especially when label noise level is high.

The detailed numerical results for each specific case contributing to the averaged values in Table~\ref{t3} and Figure~\ref{f8} are provided in the Supplementary Materials (Tables~S2-S5 and Figures~S4–S5).

\subsection{Experiments on real-world data without artificial label noise}
\begin{figure}[t]
  \centering
  \begin{tabular}{@{}c@{}}
    \includegraphics[width=1.1\textwidth, height=8cm]{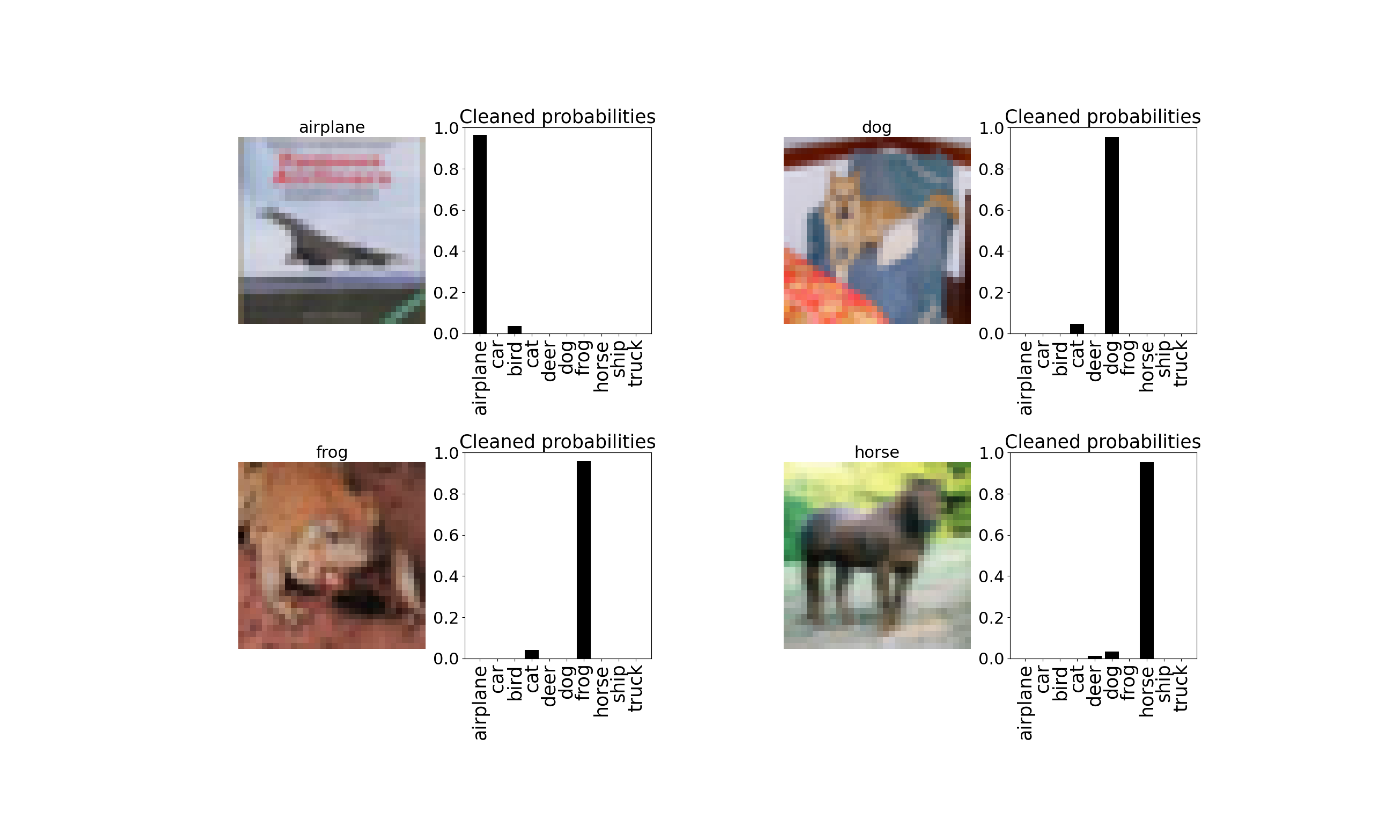}
    \\[0.001em]
    \includegraphics[width=1.1\textwidth, height=8cm]{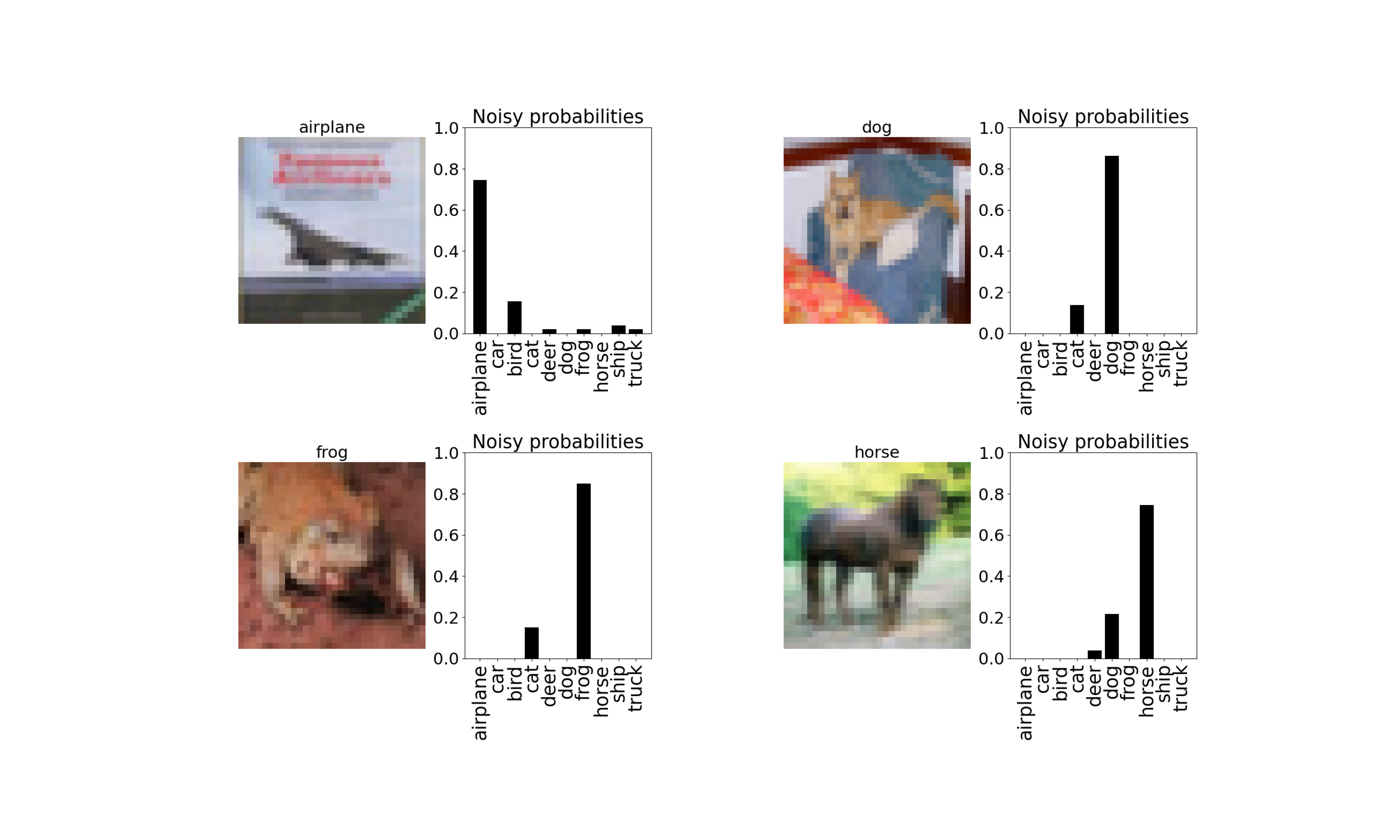}
  \end{tabular}
  \caption{Examples of CIFAR-10H soft labels. 
  Top: noisy soft label distributions. 
  Bottom: cleaned soft label distributions obtained after applying the proposed method.}
  \label{f9}
\end{figure}

We perform our study using the CIFAR-10H dataset\citep{peterson2019human}, which consists of 10000 CIFAR-10 test images annotated by multiple human annotators. Each image is associated with a soft label representing the empirical distribution of human responses, reflecting an average of 4.6\% label uncertainty. Figure~\ref{f9} shows an illustration of some noisy soft labels and how our method cleans them.

This dataset has been used in prior work as a calibration set for conformal prediction\citep{sesia2025adaptive}. Conformal prediction is a technique that produces prediction sets rather than single-label outputs. Its performance is typically evaluated using two metrics: coverage (the probability that the true label is included in the prediction set) and size (the number of labels in the prediction set). Higher coverage and lower size indicate better prediction performance.

In our experiments, we follow a similar setup: we randomly generate 20 calibration sets with sizes ranging from 500 to 9500 sampled from CIFAR-10H. A clean training set of 50,000 CIFAR-10 images is used to train the prediction model, and a random clean test set of 500 CIFAR-10 images is used for evaluation. We choose ResNet-18 convolutional neural network as the classifier. The average coverage and size over the 10 repetitions are computed for each calibration size. The results are summarized in the following tables and figures to evaluate the effectiveness of conformal prediction under different calibration set sizes.

\begin{figure}[H]
  \centering
  \begin{tabular}{@{}c@{\hspace{0.1cm}}c@{}}
    \includegraphics[width=7.7cm,height=5cm]{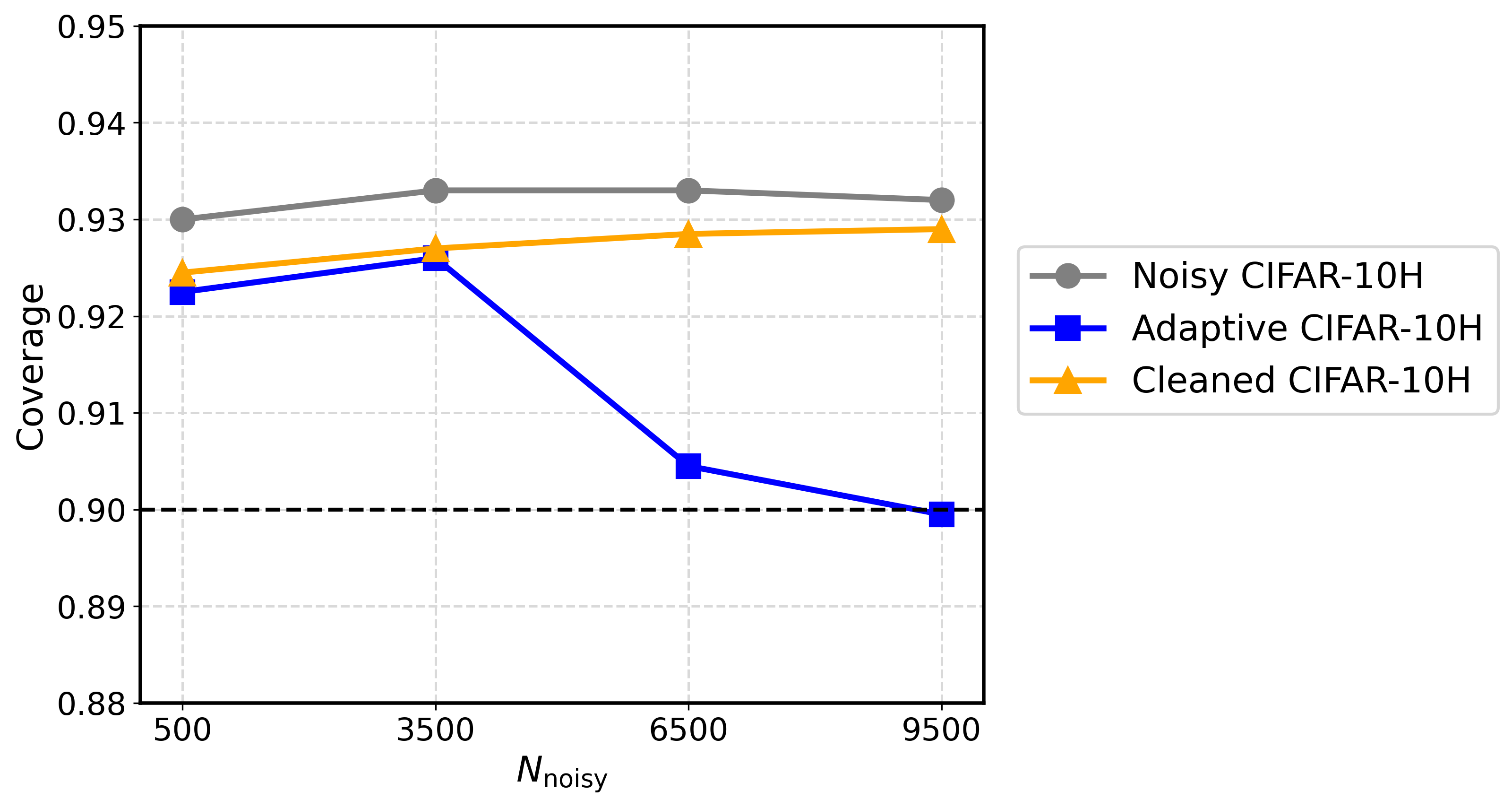} &
    \includegraphics[width=7.7cm,height=5cm]{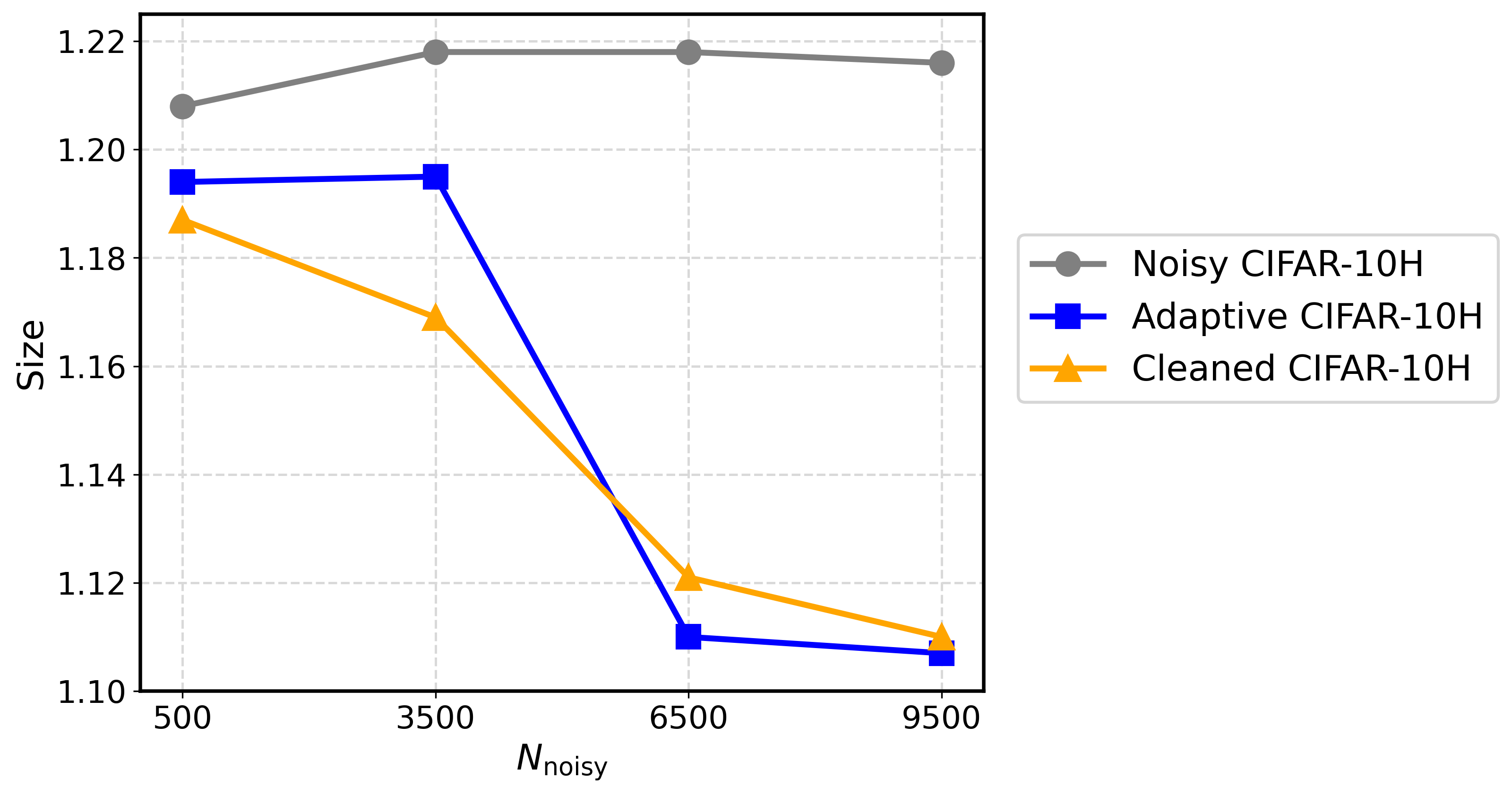}
  \end{tabular}
  \caption{Comparison between the noisy data and the adaptive method proposed by M. Sesia in terms of coverage and size, computed over 10 CIFAR-10H random calibration sets of size 500–9500.}
  \label{f10}
\end{figure}

Table~\ref{t4}  demonstrates that the ResNet-18 convolutional neural network-based label cleaning procedure on CIFAR-10H is consistent and effective. It maintains most of the clean data. Although some noisy labels remain uncorrected the overall balance between cleaning precision and data retention is strong, reducing the label noise level from 4.6\% to below 0.9\%.
\begin{table}[t]
\centering
\begin{tabular}{ccc}
\toprule
$n$ & $n_{\text{cleaned}}$ & $l_{cleaned}$ \\
\midrule
500   & 484.37  & 0.0085  \\
3500  & 3285.89 & 0.0089 \\
6500  & 6211.01 & 0.0086  \\
9500  & 8852.58 & 0.0083 \\
\bottomrule
\end{tabular}
\caption{Averaged ResNet-18 convolutional neural network label cleaning results over 20 random subsets from CIFAR-10H of sizes 500, 3500, 6500, and 9500. $n$ is the size of the noisy subset, $n_{\text{cleaned}}$ is the size after cleaning, $l_{\text{cleaned}}$ is the purified noise rate.}
\label{t4}
\end{table}
Figure~\ref{f10} illustrates the comparison between our method and the prior work in terms of coverage and prediction set size across four different calibration set sizes (500, 3500, 6500, and 9500).

The results show that our method consistently achieves higher coverage than the prior approach across all calibration sizes. This indicates that our cleaned calibration sets are more reliable in capturing the true label, making the conformal prediction more robust.

In terms of prediction set size, which reflects how informative the prediction is, our method also performs well. Specifically, it achieves smaller prediction sets than the original noisy data at all calibration sizes, confirming the benefit of removing label noise. Furthermore, our method yields smaller sizes than the prior work when the calibration size is 500 and 3500, showing both high coverage and compact predictions.

At larger calibration sizes (6500 and 9500), our method becomes slightly less informative than the prior work in terms of prediction set size. However, it still maintains a higher coverage than the nominal 90\% marginal coverage threshold, ensuring that the prediction sets remain statistically valid and reliable. These results demonstrate the overall strength of our method in balancing coverage and informativeness, especially in low and moderate size calibration settings.
\section{Conclusions}\label{sec5}

We proposed a label noise cleaning method based on Bernoulli random sampling.
First, we give full mathematical proof showing that the conditional label noise levels of Bernoulli random subsets given a clean or a noisy observation—follow two different distributions.
Because the conditional means across observations are PQD, this distinction alone does not allow direct separation of clean and noisy observations.

To address this, we introduce an independent coupling of the conditional label noise levels.
By proving that the conditional means converge to this independent coupling, we can apply SLLN, which implies that the empirical distribution of the conditional label noise levels converges to a two-component mixture of the two distinct distributions mentioned above.

Then, building on a well-supported linear model between label noise level and averaged cross-validation error, we approximate this empirical distribution.
Because the two distributions are distinct, their midpoint serves as a natural cut point for separating clean and noisy observations.
Moreover, since the midpoint divides the density into equal mass on both sides, we can estimate this cut point directly from the averaged cross-validation errors, without estimating any parameters or requiring prior knowledge of which labels are correct.

Experiments on both simulated and real datasets show that this method achieves strong noise reduction and improved classification accuracy.
Across different settings, it performs well on multiple measures of label noise cleaning, including the residual label noise level in the cleaned data, the proportion of clean samples retained, the proportion of noisy samples removed, and the classification performance under various supervised classifiers.
Overall, the results are consistently better or comparable to existing label cleaning studies. 
\section*{Supplementary Material}
The supplementary material contains the proofs and additional simulation results.
\bibliography{reference}

\end{document}